\pdfoutput=1
\documentclass[aps,prb,floatfix,twocolumn,superscriptaddress,showpacs,amsmath,amssymb]{revtex4-1}
\usepackage{graphicx}
\usepackage{epstopdf}
\usepackage{epsfig}
\usepackage{epsf}
\usepackage{url}
\usepackage[USenglish]{babel}

\usepackage{hyperref}
\usepackage{babel}

\allowdisplaybreaks

\renewcommand\[{\begin{equation}}
\renewcommand\]{\end{equation}}

\begin{document}
\title{Embedding Dynamical Mean-Field Theory for Superconductivity in Layered Materials and Heterostructures}
\author{Francesco Petocchi}
\affiliation{International School for Advanced Studies (SISSA/ISAS) and CNR-IOM Democritos, Via Bonomea 265, 34136, Trieste, Italy}
\author{Massimo Capone}
\affiliation{International School for Advanced Studies (SISSA/ISAS) and CNR-IOM Democritos, Via Bonomea 265, 34136, Trieste, Italy}
\begin{abstract}
We study layered systems and heterostructures of s-wave superconductors by means of a suitable generalization of Dynamical Mean-Field Theory. In order to reduce the computational effort, we consider an embedding scheme in which a relatively small number of active layers is embedded in an effective potential accounting for the effect of the rest of the system. We introduce a feedback of the active layers on the embedding potential that improves on previous approaches and essentially eliminates the effects of the finiteness of the active slab allowing for cheap computation of very large systems.
We extend the method to the superconducting state, and we benchmark the approach by means of simple paradigmatic examples showing some examples on how an interface affects the superconducting properties. As examples, we show that superconductivity can penetrate from an intermediate coupling superconductor into a weaker coupling one for around ten layers, and that the first two layers of a system with repulsive interaction can turn superconducting by proximity effects even when charge redistribution is inhibited.
\end{abstract}
\maketitle
\section{Introduction}
The advances in manufacturing and handling heterostructures are in the forefront of solid state research. In particular heterostructure based on oxides have a huge potential thanks to the rich physics of their constituents. Combining different oxides one can even engineer and tailor electronic and magnetic states which can be completely different from those of the bulk constituents. 
The possibility to control these emergent and intrinsic properties of the constituents opens an avenue towards the realization of new devices based on correlated electrons. 

One of the most studied examples of the novel physics at oxide interfaces is the appearance of a high-mobility electron gas at the interface between the
band insulator $\mathrm{SrTiO_{3}}$ (STO) and the Mott insulator $\mathrm{LaTiO_{3}}$ (LTO)\cite{Ohtomo2002}. This nearly two-dimensional metal can be   easily manipulated through gate voltages and turned into a superconductor\cite{Biscaras2010,Biscaras2012} which strikingly appears combining two non-superconducting materials. Superconductivity has also been observed at interfaces between two band insulators such as STO and LaAlO$_3$\cite{Reyren2007}, while interfaces between different copper-based superconductors have a critical temperature higher than the bulk constituents\cite{Hwang2012,Chakhalian2014}. 
These are just examples of a variety of phenomena involving superconductivity in artificially crafted heterostructures. The aim of this work is to develop a reliable formalism to study superconductivity in heterostructures beyond simplifying limits such as the Bardeen-Cooper-Schrieffer (BCS) approximation. In order to test the method and to single out the effect of intermediate- and strong-coupling, we consider a simple attractive Hubbard model as a simple paradigmatic model for an s-wave superconductor.

The theoretical description of interacting heterostructures requires methods which are at the same time able to treat the relevant interactions and to effectively account for the geometrical arrangement of these systems. The intrinsic difficulty to solve interacting systems beyond the perturbative regime limits the number of accessible approaches and prompts for the use of suitable approximations. Dynamical Mean-Field Theory (DMFT) has indeed demonstrated to accurately treat the competing interactions characterizing oxide interfaces including electron-electron interactions\cite{Georges1996}, electron-phonon coupling\cite{Freericks1993,Millis1996,Meyer2002,Capone2003} and their interplay\cite{Freericks1995,Deppeler2002,Koller2004,Sangiovanni2005,Werner2007,Sangiovanni2006,Capone2010,Giovannetti2014,Nomura2015} as well as for the attractive Hubbard model\cite{Keller2001,Capone2002,Toschi2005a,Toschi2005b,Garg2005,Bauer2009a,Bauer2009b,Koga2011}.

The extension of DMFT to treat surface and interface effects has been pioneered by   Potthoff and Nolting\cite{Potthoff1999a,Potthoff1999b} who introduced a layer generalization of DMFT and applied it to a solid-vacuum interface in the presence of short-range Coulomb interaction as described by the Hubbard model. Including also long-range Coulomb interactions Okamoto and Millis\cite{Okamoto2004a,Okamoto2004b} and Kancharla and Dagotto\cite{Kancharla2006} have considered charge-transfer effects and proposed that the charge leakage from one layer to another is responsible for the metallic interface between LTO and STO. 
As we discuss in more details in the following, when DMFT is extended to inhomogeneous systems, the inclusion of more and more layers is the bottleneck of the calculation. Therefore the main limitation of these approaches is the influence of finite-size effects and the slow convergence to the bulk limit (infinite number of layers). Ishida and Liebsch have proposed and implemented\cite{Ishida2009,Ishida2010} a strategy to overcome this limit.
The idea is to effectively describe a substrate with an energy-dependent embedding potential. 

In this work we extend the embedding potential to study superconductivity with s-wave symmetry within each layer and we introduce a ``feedback" effect which improves the performance of the embedding method. The purpose of this paper is to demonstrate the feasibility of this approach for superconducting state and to study the evolution of the physics as a function of the coupling strength. The extension to cluster methods, which is necessary to study d-wave superconductivity is conceptually simple but computationally demanding. 

The paper is organized as follows. In Sec. II we introduce the model and the general concept of layered DMFT and the embedding approach. Sec. III is dedicated to the extension of the approach to superconducting systems and to our recipe for the embedding potential. Sec. IV  describes our results for different physical configurations, while Sec. V contains conclusions and perspectives.

\section{Model and Method}
In this section we introduce our approach to extend previous DMFT-based approaches to heterostructure to allow for superconductivity. For the sake of clarity, in the first subsection we briefly review the DMFT formalism in the superconducting state and some aspects of the  exact diagonalization (ED) solution of DMFT that we employ in our practical implementation. 
\subsection{Single-site DMFT and superconductivity}
Dynamical Mean-Field Theory is one of the most popular and successful theoretical methods to treat strongly correlated electron systems. It extends the classical mean-field approach to the quantum dynamical domain by mapping a lattice model onto an impurity model in which an interacting lattice site is hybridized with a non-interacting bath which is self-consistently determined. 

In this section we present the DMFT formalism for superconducting solutions starting from the attractive Hubbard model, which can be considered the simplest model for an s-wave superconductor. However the same equation would be found for example for an electron-phonon model or even for models without an explicit source of pairing. 

The Hamiltonian reads
\begin{eqnarray}
\label{hubbard}
{\cal H} &= &-t \sum_{<ij>\sigma} (c_{i\sigma}^{\dagger} c_{j\sigma} +\mathrm{h.c.})
-\sum_i \mu (n_{i\uparrow}+n_{i\downarrow})
\nonumber\\
& &- U\sum_{i} n_{i\uparrow}n_{i\downarrow}, 
\end{eqnarray}
where the sums run over the sites $i$ and $j$ of a lattice, $c_{i\sigma} (c^{\dagger}_{i\sigma})$ are annihilation (creator) operators for fermions with spin $\sigma$ on site $i$, $t$ is a nearest-neighbor hopping amplitude, $U$ is a positive energy measuring the strength of the on-site attractive interaction and $\mu$ is the chemical potential. This model is known to have an s-wave superconducting ground state for any value of the coupling $U$ and it has been extensively studied by means of DMFT\cite{Keller2001,Capone2002,Toschi2005a,Toschi2005b,Garg2005,Bauer2009a,Bauer2009b,Koga2011}.

As mentioned above, within DMFT the lattice model is mapped onto an impurity model which, for an attractive Hubbard model, may be written as
\begin{equation}
\begin{array}{c}
\mathcal{H}_{imp}=\sum_{l\sigma}^{n_{s}}\left[\varepsilon_{l}\hat{c}_{l\sigma}^{\dagger}\hat{c}_{l\sigma}+V_{l}\left(\hat{c}_{l\sigma}^{\dagger}\hat{d}_{o\sigma}+h.c.\right)\right]\\
\left.+\frac{\Delta_{l}}{2}\left(\hat{c}_{l\sigma}\hat{c}_{l\bar{\sigma}}+h.c.\right)\right]-U\hat{n}_{o\uparrow}\hat{n}_{o\downarrow}-\mu\left(\hat{n}_{o\uparrow}+\hat{n}_{o\downarrow}\right)
\end{array}\label{eq:SC_imp_Ham}
\end{equation}
where $\hat{c}_{l\sigma}^{\dagger}$
creates a particle in $l-th$ level of a non-interacting bath which is parameterized by the energy levels $\varepsilon_l$ and the superconducting amplitudes $\Delta_l$ and by the hybridizations $V_l$. The amplitudes $\Delta_l$ give rise to an anomalous (superconducting) component of the hybridization function between the impurity and the bath which is necessary to treat the superconducting phase. 
Solving the impurity model and computing the normal and anomalous Green's functions $G=-\left\langle T{c}\left(\tau\right){c}^{\dagger}\right\rangle $
and $F=-\left\langle T{c}^{\dagger}\left(\tau\right){c}^{\dagger}\right\rangle $,
we can obtain the impurity self-energy as $\hat{\Sigma}_{imp}=\hat{\mathcal{G}}_{o}^{-1}-\hat{G}_{imp}^{-1}$, where the hat denotes 2$\times$ 2 matrices whose components are given by the normal and anomalous Greens' functions
\begin{eqnarray}
\hat{G}_{\alpha}=\left(\begin{array}{cc}
G\left(i\omega_{n}\right) & F\left(i\omega_{n}\right)\\
F\left(i\omega_{n}\right) & -G\left(i\omega_{n}\right)^{\star}
\end{array}\right)
\end{eqnarray}
and analogously for the two components of the ``Weiss field", which coincide with the non-interacting Green's functions of (\ref{eq:SC_imp_Ham}).
$\hat{\mathcal{G}}_{o}$:
\begin{equation}
\begin{array}{c}
\mathcal{G}_{o\left(11\right)}^{-1}\left(i\omega_{n}\right)=i\omega_{n}+\mu-\sum_{l}^{n_{s}}\left|V_{l}\right|^{2}\frac{i\omega_{n}+\varepsilon_{l}}{\omega_{n}^{2}+\varepsilon_{l}^{2}+\Delta_{l}^{2}}\\
\mathcal{F}_{o\left(12\right)}^{-1}\left(i\omega_{n}\right)=\sum_{l}^{n_{s}}\left|V_{l}\right|^{2}\frac{\Delta_{l}}{\omega_{n}^{2}+\varepsilon_{l}^{2}+\Delta_{l}^{2}}.
\end{array}\label{eq:Weiss_fields}
\end{equation}
The DMFT approximation is enforced requiring that the local Green's functions defined above coincide with the local components of the lattice Green's function $\hat{G}_{lat}\left(i\omega_{n}\right)=\int d\varepsilon\rho\left(\varepsilon\right)\left[i\omega_{n}\mathbf{1}^{\left(2\right)}+\left(\mu-\varepsilon\right)\hat{\sigma}_{3}-\hat{\Sigma}_{imp}\left(i\omega_{n}\right)\right]^{-1}$, being $\rho\left(\varepsilon\right)$ the non interacting density of states. 

A practical implementation of DMFT requires to recursively solve the impurity model calculating $G$ and $F$. This allows to compute the self-energy matrix and a new  Weiss field $\hat{\mathcal{G}}_{o}^{-1}=\hat{\Sigma}_{imp}+\hat{G}_{lat}^{-1}$.  The process is iterated until the Weiss fields and the other quantities are converged. A central issue in DMFT calculation is indeed the solution of the impurity model. Here we use an exact diagonalization ``solver", in which the groundstate of the impurity Hamiltonian is found using a Lanczos algorithm. In order to obtain a finite matrix, the sums over $l$ are truncated to a finite and small value $N_b$. Nonetheless, small values of $N_b$ have been shown to be sufficient to obtain converged results for thermodynamic observables. The ED solution of DMFT involves one more step with respect to the algorithm we described. After the self-consistency condition is used to find new Weiss field, these functions need to be cast in the form (\ref{eq:Weiss_fields}) with a discrete value of $N_b$. This can be achieved by fitting the new Weiss fields with Eq. (\ref{eq:Weiss_fields}) which has to be interpreted as a function of the ``Anderson parameters" $\varepsilon_l$, $V_l$ and $\Delta_l$. 

\subsection{Observables}
To characterize the superconducting states of our layered superconductor and its spatial dependence we use several observables. 
The most direct evidence of the superconducting state and its strength is the layer-resolved zero-temperature pairing amplitude, simply obtained  as the integral of the anomalous part of the $\alpha-th$
layer Green' function 
\begin{equation}
\Delta_{\alpha}=T\sum_{n}F_{\alpha}\left(i\omega_{n}\right).
\end{equation}
The nature of the superconducting state (for example if the system is in an effective weak- or strong-coupling regime) can be characterize 
in terms of the different contribution to the total energy. The layer-resolved potential energy is simply 
\begin{equation}
E_{pot}^{\alpha}=U \left\langle \hat{n}_{o\uparrow}\hat{n}_{o\downarrow}\right\rangle _{\alpha},
\end{equation}
while the kinetic energy reads:
\begin{equation}
\left\langle E_{k}^{\alpha}\right\rangle =T\sum_{n}\intop d\epsilon\rho\left(\epsilon\right)Tr\left\{ \epsilon\hat{\sigma}_{3}\hat{G}_{\alpha}\left(\epsilon,i\omega_{n}\right)\right\}.
\end{equation}
Notice that while the global order parameter and potential energy are simply obtained by summing the contributions from the different layers, the bulk kinetic energy also includes the contributions from the interlayer hoppings, which do not contribute to the above $\langle E_k\rangle$. 
Finally we can compute the  quasparticle weight,
namely $z_{\alpha}=\left(1-\partial\Sigma_{\alpha}^{11}\left(i\omega_{n}\right)/\partial\left(i\omega_{n}\right)\right)^{-1}$.
Where $\Sigma_{\alpha}^{11}\left(i\omega_{n}\right)$ is the normal
component of the $\alpha-th$ layer self-energy, which measures the coherence of the low-energy excitations.

\subsection{Superconducting DMFT applied to heterostructures and embedding potentials}
In the previous subsections we introduced single-site DMFT for bulk superconductors, in which full translational invariance is enforced and any lattice site is equivalent. In order to study layered systems we need to use a suitable extension of DMFT able to treat inhomogeneous systems with a layered geometry.
We focus on a simple cubic lattice partitioned into $N$ layers  stacked along the (001) direction. Within each layer translational invariance is assumed and the  two-dimensional wavevector $\mathbf{k}_{\Vert}=\left(k_{x},k_{y}\right)$ is a conserved quantity.

The Green's function $\hat{G}_{S}$ of a slab made of $N$ superconducting
layers can be expressed as a $2N\times2N$ matrix corresponding to the two components of Nambu spinors and to the $N$ layers
\begin{equation}
\bar{G}_{S}\left(\mathbf{k},i\omega_{n}\right)=\left[\left(i\omega_{n}\mathbf{1}^{\left(2\right)}+\mu\hat{\sigma}_{3}\right)\otimes\mathbf{1}^{\left(N\right)}-\bar{\varepsilon}_{\mathbf{k}}-\bar{\Sigma}\left(i\omega_{n}\right)\right]^{-1}\label{eq:layer_GF_no_emb}
\end{equation}
where $\mathbf{1}^{\left(N\right)}$ is the $N$-dimensional identity
matrix and $\hat{\sigma}_{3}$ is the third Pauli matrix. 2N$\times$2N matrices are identified by a bar.
The single-particle dispersion matrix is given by:
\begin{eqnarray}
\bar{\varepsilon}_{\mathbf{k}}=\left(\begin{array}{ccc}
\varepsilon_{1}^{\Vert}\hat{\sigma}_{3} & \varepsilon_{12}^{\bot}\hat{\sigma}_{3} & 0\\
\varepsilon_{21}^{\bot}\hat{\sigma}_{3} & ... & ...\\
0 & ... & \varepsilon_{N}^{\Vert}\hat{\sigma}_{3}
\end{array}\right)
\end{eqnarray}
whose elements are: $\varepsilon_{\alpha\alpha}^{\Vert}=-2t_{\alpha}\left[\cos\left(k_{x}\right)+\cos\left(k_{y}\right)\right]$
and $\varepsilon_{\alpha\beta}^{\bot}=t_{z}$. The self-energy matrix is instead a block-diagonal matrix 
\begin{equation}
\bar{\Sigma}\left(i\omega_{n}\right)=\left(\begin{array}{ccc}
\hat{\Sigma}_{1} & 0 & 0\\
0 & ... & 0\\
0 & 0 & \hat{\Sigma}_{N}
\end{array}\right)\label{eq:Self-energy_matrix}
\end{equation}
where each element is a $2\times2$ block with normal and anomalous
components corresponding to the local self-energy of an individual layer. The underlying approximation is that the self-energy remains local $\bar{\Sigma}_{ij} = \delta_{ij}\bar{\Sigma}_i$ and it is uniform within each layer, while the different layers are allowed to have different self-energies $\bar\Sigma_{\alpha}$, each associated to a local effective impurity problem. An explicit solution requires to solve as many impurity models as the number of layers, from which the individual self-energies are obtained and plugged into Eq. (\ref{eq:layer_GF_no_emb}). Summing over the momenta within each layer leads to a set of local Green's functions which are then imposed to coincide with the impurity Green's functions.

As opposed to single-site DMFT, we are therefore limited to a finite system along the z direction, which can lead to  finite-size effects, that are enhanced if we use an open slab, where the electrons on the outmost layers become effectively more interacting because of the missing neighbors ( Fig.(\ref{fig:1})). A possible solution to overcome this limitation is to sandwich the finite slab of N layers into two media\cite{Ishida2009,Ishida2010} effectively accounting for the presence of bulk layers.

Using the notations of Ref.  \cite{Nourafkan2009} we define the matrix inverse of (\ref{eq:layer_GF_no_emb})
\begin{equation}
\hat{A}_{S}\left(\mathbf{k},i\omega_{n}\right)\hat{G}_{S}\left(\mathbf{k},i\omega_{n}\right)=\mathbf{1}^{\left(2N\right)}\label{eq:inverse}
\end{equation}
Partitioning the infinite three-dimensional system into a slab (S) and two ``bulk" samples (B$_R$ and B$_L$) we can rewrite Eq. (\ref{eq:inverse}) as \begin{equation}
\left(\begin{array}{ccc}
\hat{A}_{B_{L}} & \hat{A}_{B_{L}1} & 0\\
\hat{A}_{1B_{L}} & \hat{A}_{S} & \hat{A}_{B_{R}N}\\
0 & \hat{A}_{NB_{R}} & \hat{A}_{B_{R}}
\end{array}\right)\left(\begin{array}{ccc}
\hat{G}_{B_{L}} & \hat{G}_{B_{L}1} & 0\\
\hat{G}_{1B_{L}} & \hat{G}_{S} & \hat{G}_{B_{R}N}\\
0 & \hat{G}_{NB_{R}} & \hat{G}_{B_{R}}
\end{array}\right)=\mathbf{1}\label{eq:general_matrix}
\end{equation}
where the matrices are now $\left(2N+4\right) \times \left(2N+4\right) $. The diagonal elements of the two matrices are the Green's functions and the $A$ functions for the slab (S) and the two semi-infinite substrate which embed the interacting slab. The non-zero off-diagonal elements describe the processes connecting the ``left" efffective substrate with layer 1 of the slab and the ``right" substrate with layer N of the slab. From this we can single out the equation for the slab Green's function
\begin{equation}
\left(\hat{A}_{S}-\hat{A}_{1B_{L}}\hat{A}_{B_{L}}^{-1}\hat{A}_{B_{L}1}-\hat{A}_{NB_{R}}\hat{A}_{B_{R}}^{-1}\hat{A}_{B_{R}N}\right)\hat{G}_{S}=\mathbf{1}^{\left(2N\right)}
\end{equation}
which shows how the 1 and N indices are affected directly by the presence of the two semi-infinite bulks. The explicit result is 
\begin{equation}
\begin{array}{c}
\hat{G}_{S}\left(\mathbf{k},i\omega_{n}\right)=\left[\left(i\omega_{n}\mathbf{1}^{\left(2\right)}+\mu\hat{\sigma}_{3}\right)\otimes\mathbf{1}^{\left(N\right)}-\hat{\varepsilon}_{\mathbf{k}}\right.\\
\left.-\hat{\Sigma}\left(i\omega_{n}\right)-\delta_{\alpha1}\hat{S}_{B_{L}}\left(\mathbf{k},i\omega_{n}\right)-\delta_{\alpha N}\hat{S}_{B_{R}}\left(\mathbf{k},i\omega_{n}\right)\right]^{-1}
\end{array}\label{eq:layer_Gf_with_emb}
\end{equation}
where we have defined the complex embedding potentials 
\begin{equation}
\hat{S}_{B_{L,R}}\left(\mathbf{k},i\omega_{n}\right)=t_{z}^{2}\hat{\sigma}_{3}\hat{G}_{B_{L,R}}\left(\mathbf{k},i\omega_{n}\right)\hat{\sigma}_{3}\label{Eq:potential}
\end{equation}
\emph{acting on the first and last layer only} if the interlayer hopping is limited to nearest neighbors.  Comparing with  (\ref{eq:layer_GF_no_emb}) it is evident that the only difference is introduced by the embedding potential at the boundaries of the slab. 
\subsection{Choice of the substrate Green's functions}

Here we introduce an optimized strategy to describe heterostructures in terms of a few ``active" layers embedded between two semi-infinite systems. The starting point is naturally the surface Green's function of a semi-infinite system. We partition a bulk system in two semi-infinite halves along the direction of the layers of the heterostructure\cite{Kalkenstein1971}. As in the rest of this work, the Green's functions are assumed to be translational invariant along each layer and they are labelled according to the layer index. We denote the surface layer with 0 and, for the sake of definiteness, we focus on the ''left'' system with negative layer index. 

The relation between the Green function of the $\alpha-th$ layer
in the left semi-infinite bulk $\hat{G}_{\alpha}$ and the
same layer in the bulk crystal $\hat{G}_{\alpha}^{bulk}$ can be written as
\begin{equation}
\hat{G}_{\alpha}=\hat{G}_{\alpha}^{bulk}+\sum_{\beta}\hat{G}_{\alpha}^{bulk}\hat{V}_{\alpha\beta}\hat{G}_{\beta}\label{eq:Dyson_for_Gf_bulk},
\end{equation}
where $V_{\alpha\beta}$ are the hopping matrix elements connecting the right and left sides with indexes  $\alpha\in0,-1,-2,\ldots$ and $\beta =1, 2,\ldots$. We are only interested in the surface layer 0 which is connected only with the next layer 1 by the diagonal (in the Nambu space) hopping matrix $\hat{T}_{01}$, which leads to 
\begin{equation}
\hat{G}_{0}=\hat{G}_{0}^{bulk}\left(\mathbf{1}^{\left(2\right)}-\hat{G}_{1}^{bulk}\hat{T}_{01}\right)^{-1}\label{eq:G_leads_no_feed},
\end{equation}
which requires the knowledge of the bulk Green's function for the surface layer and for the first layer on the left, which can easily be computed within DMFT as

\begin{equation}
\hat{G}_{\alpha}^{bulk}=\int\frac{dk}{2\pi}\frac{e^{ik\alpha}}{i\omega_{n}\mathbf{1}^{\left(2\right)}+\left(\mu-\varepsilon^{\Vert}-2t_{z}\cos k\right)\hat{\sigma}_{3}-\hat{\Sigma}\left(i\omega_{n}\right)},
\label{eq:G_formula}
\end{equation}
where the Green's function depend on the momenta along the layers $\mathbf{k}_{\Vert}$ while the integral in the LHS is performed over the transverse momentum. The self-energy in Eq. (\ref{eq:G_formula}) is determined self-consistently solving two more impurity models coupled with the slab.  The $\hat{G}_{0}$ in Eq. (\ref{eq:G_leads_no_feed}) defines the left embedding potential in Eq.  (\ref{eq:layer_Gf_with_emb}). The right potential is obviously identical.

\begin{figure}
  \includegraphics[width=0.44\textwidth]{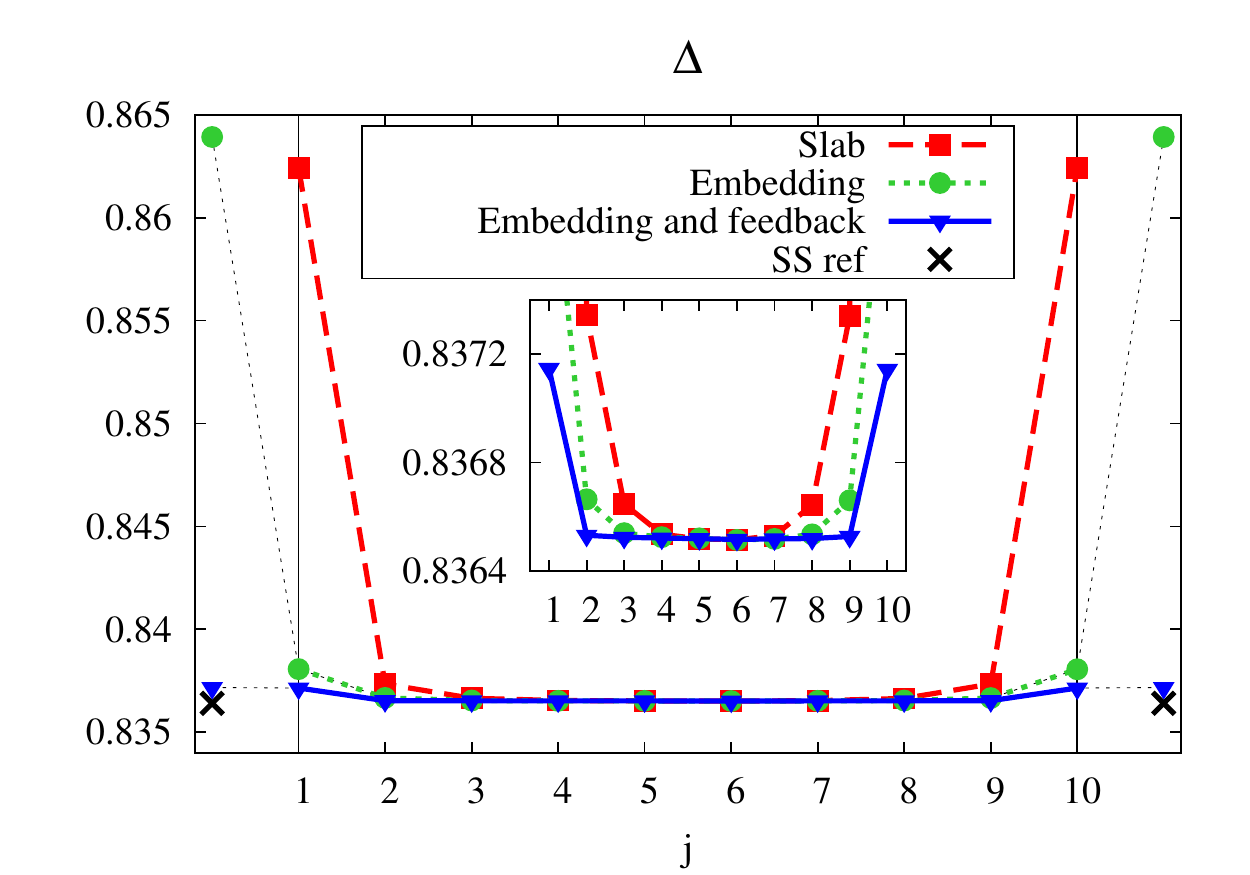}
  \caption{(Color online) Validation of out embedding scheme with feedback.
    We plot $\Delta$ for a $10$-layer embedded system with
uniform $U/t=-9$ at half-filling. The red line with squares is for the open slab, the green with dots marks the
results with the embedding potential, while the blue line with triangles denotes
the data corrected with the feedback of the slab. Black crosses on the two sides of the slab
report the bulk DMFT result. The embedding+feedback results are essentially uniform.\label{fig:1}}
\end{figure}

Fig. \ref{fig:1} presents results for a ten-layer slab for uniform parameters $U/t =-9$ and half-filling. In the absence of any embedding potential,
the slab breaks translational symmetry and the order parameter $\Delta$ becomes larger at the edges. Introducing the embedding potential according to the described scheme, we obtain the results shown as a dotted green line with large dots in Fig. \ref{fig:1}. Here we consider completely uniform parameters, and the embedding 
potential strongly reduces the inhomogeneity, even if a minor enhancement of the order parameter is clear at the edges of the slab.

In order to further reduce the effects of the finiteness of the slab,  in this work we propose a simple strategy to improve the scheme, introducing a {\it{feedback}} of the slab on the semi-infinite bulks. The idea is simply to define a potential created by the slab onto the semi-infinite bulks on the two edges. As a matter of fact the equation amounts to add a potential of the form (\ref{Eq:potential}) to the self-energy of each semi-infinite systems 
\begin{equation}
\hat{S}_{{FB}_{L,R}}\left(\mathbf{k},i\omega_{n}\right)=t_{z}^{2}\hat{\sigma}_{3}\hat{G}_{S_{1,N}}\left(\mathbf{k},i\omega_{n}\right)\hat{\sigma}_{3}.
\end{equation}

The data in Fig.  (\ref{fig:1}) demonstrate that the feedback further reduces the inhomogeneity and it
allows to essentially reproduce the uniform bulk even with a very limited number of layers.

\begin{figure}
\includegraphics[scale=0.35]{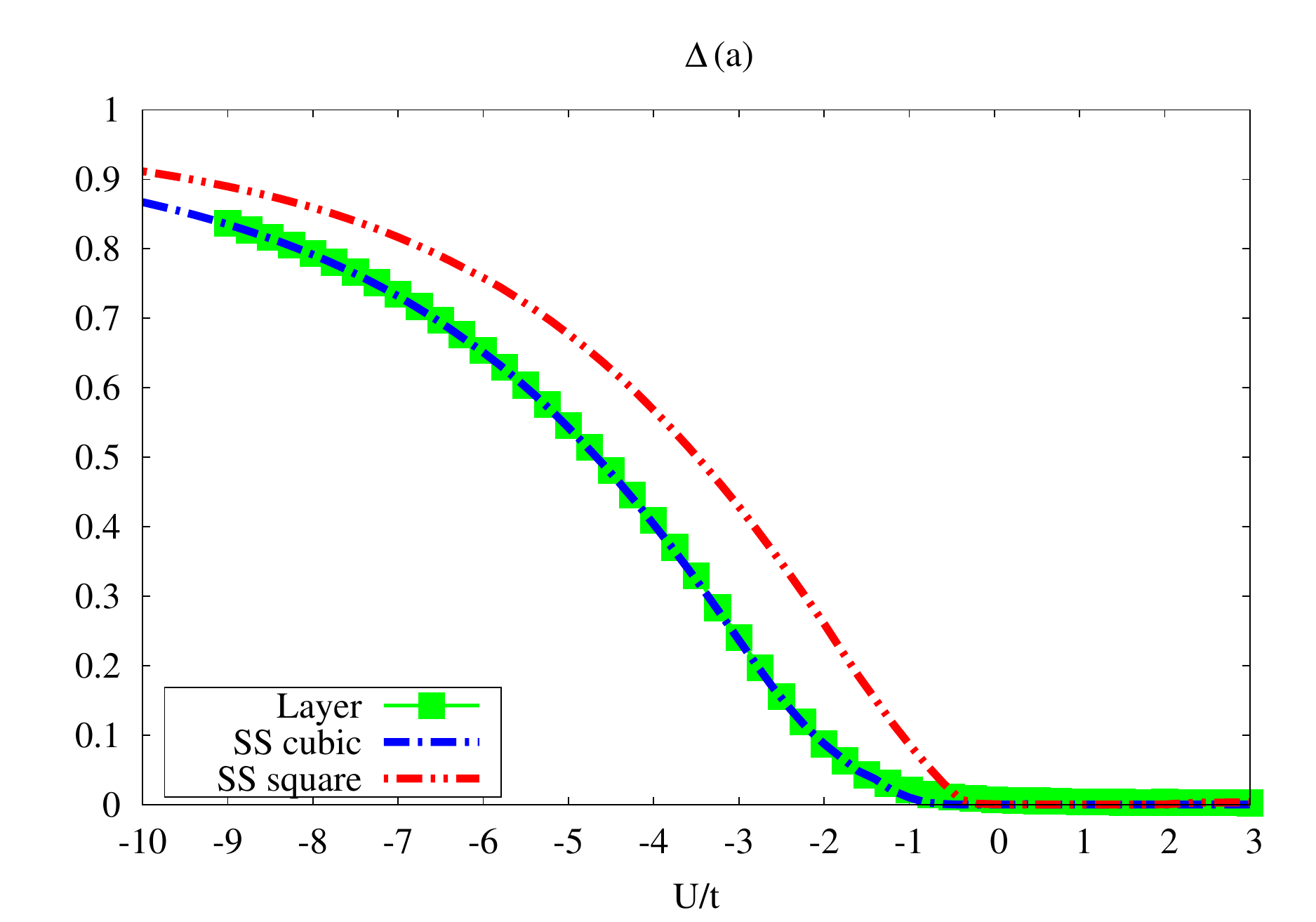}
\includegraphics[scale=0.35]{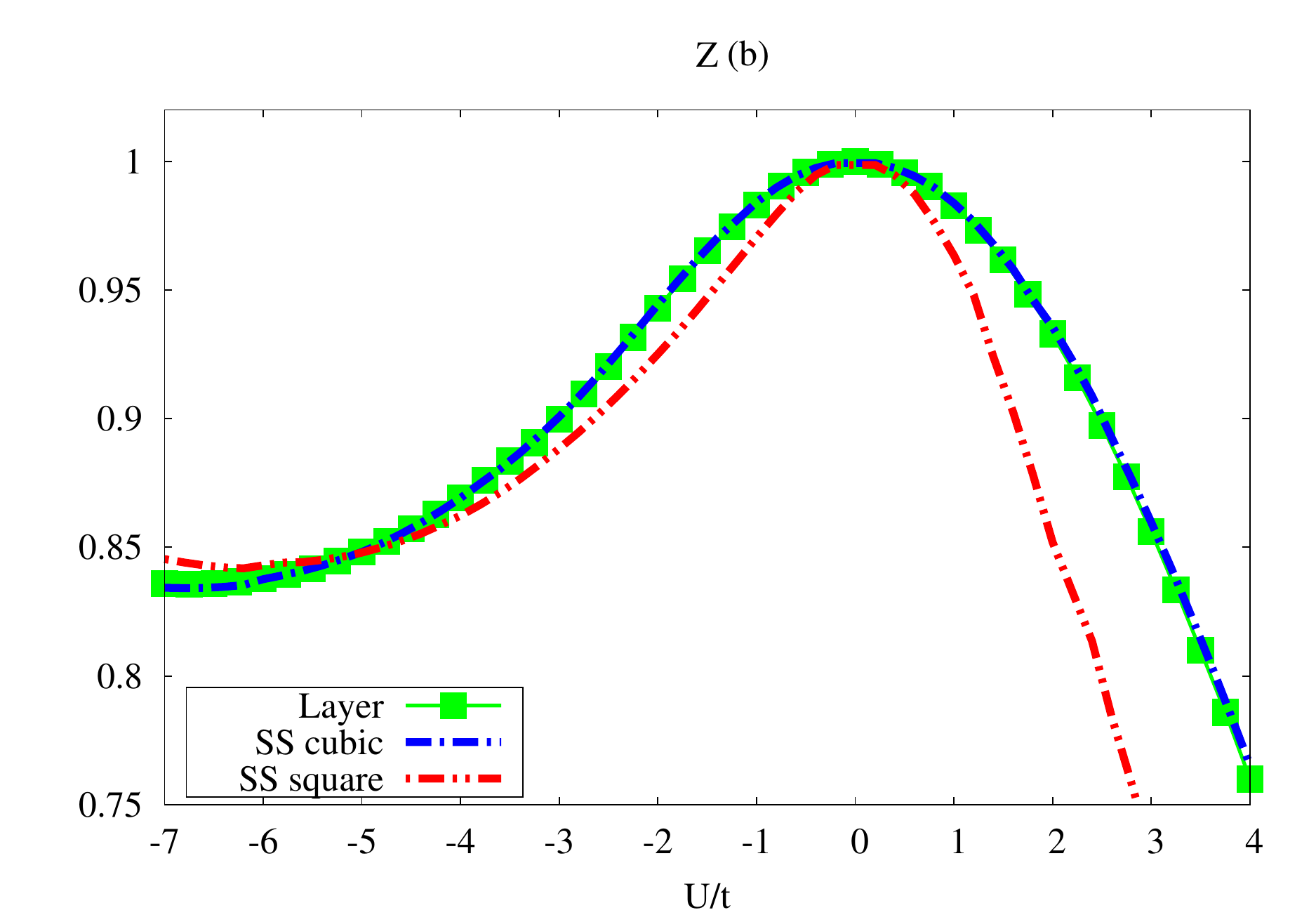}
\includegraphics[scale=0.35]{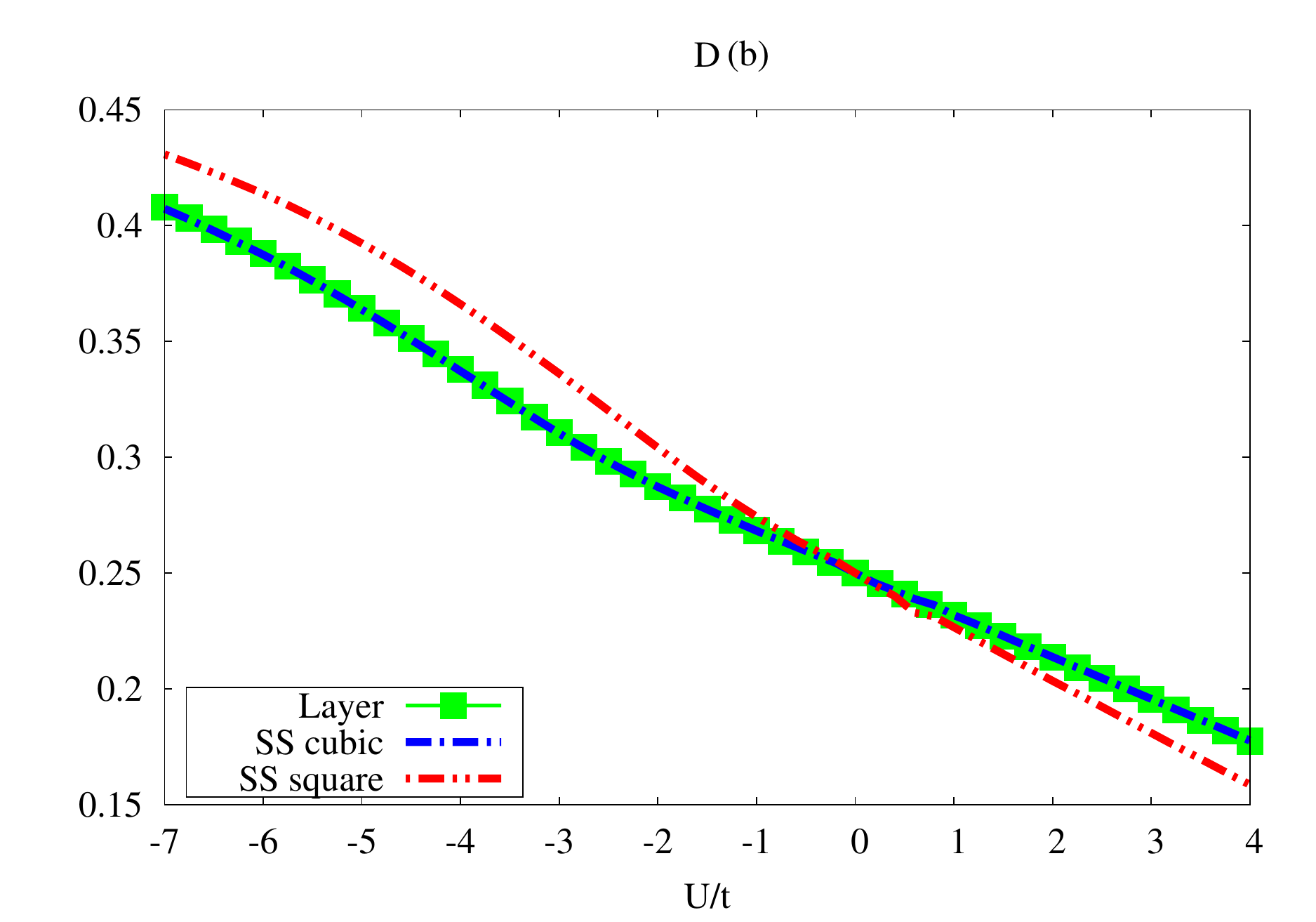}
\caption{(a) Order parameter $\Delta$, (b) quasiparticle weight $z$, (c)
double occupations expectation value $D=\left\langle \hat{n}_{o\uparrow}\hat{n}_{o\downarrow}\right\rangle $
vs interaction strength $U$ at half-filling. Red and blue lines refers
respectively to the square and cubic lattice, single site DMFT calculations.
Green dots represents the results for the central plane of an homogeneous
system made of $7$ layers, similar to the blue one in Fig.(\ref{fig:1}).
\label{fig:2}}
\end{figure}

In Fig. \ref{fig:2} we demonstrate that our feedback performs accurately for different observables and for any value of the parameters.
Here we plot the average over the slab of $\Delta$, $Z$ and of the double occupancy $D$ as a function of $U$ and we compare 
with a bulk cubic lattice (which should be reproduced when the finite-slab effect are canceled) and, for reference with a two-dimensional
calculation corresponding to a single layer. To illustrate the general validity of our approach we consider both a negative $U$, for which 
we find superconductivity, and a positive $U$ model, in which s-wave superconductivity can not establish and therefore represents 
the normal state.  The three panels of Fig. \ref{fig:2} clearly show that for every value of $U$ the three observables coincide with their
bulk counterparts.

\section{Results}

\subsubsection{Weak/strong interacting superconductor}
\begin{figure} 
\includegraphics[scale=0.36]{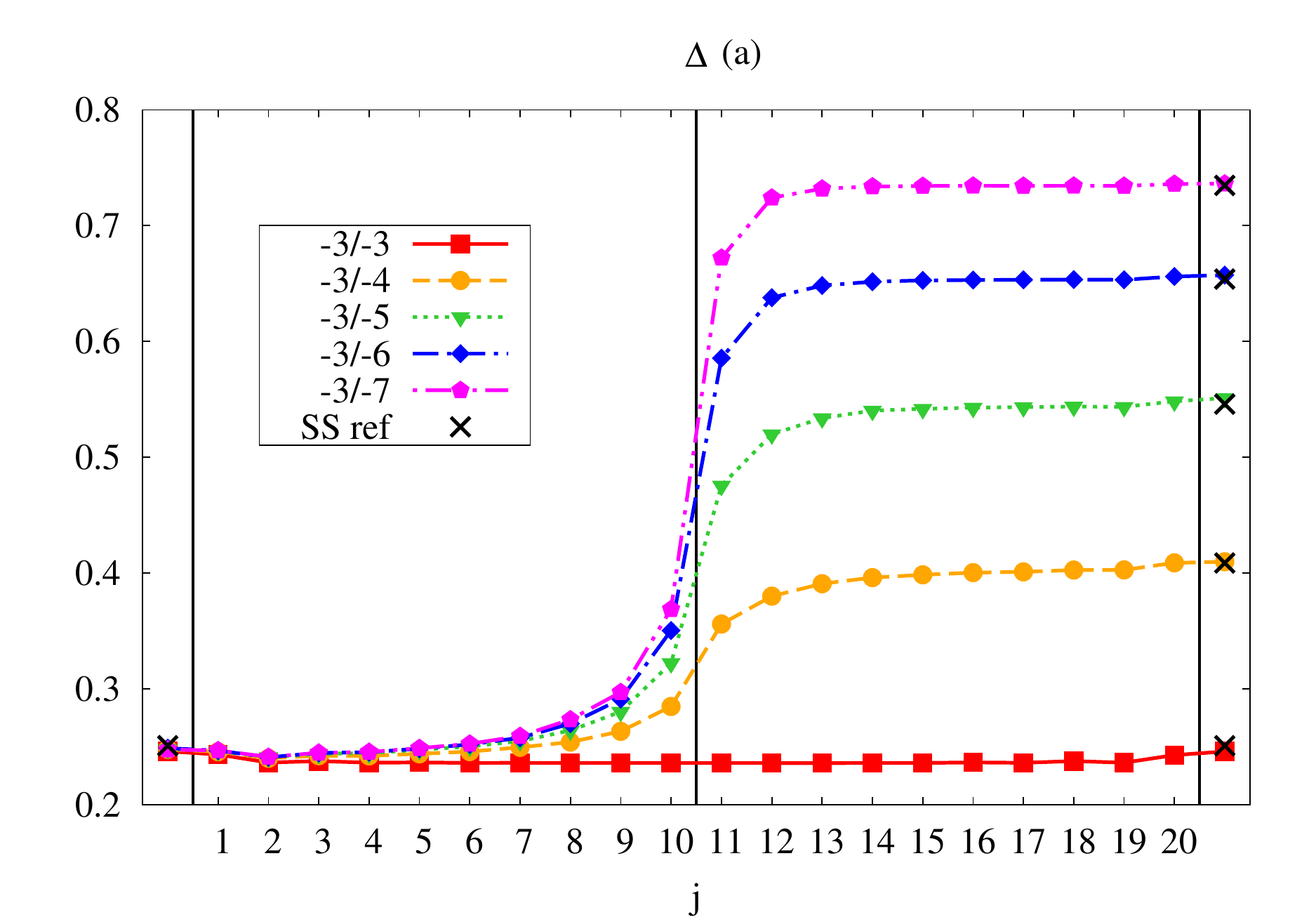}
\includegraphics[scale=0.36]{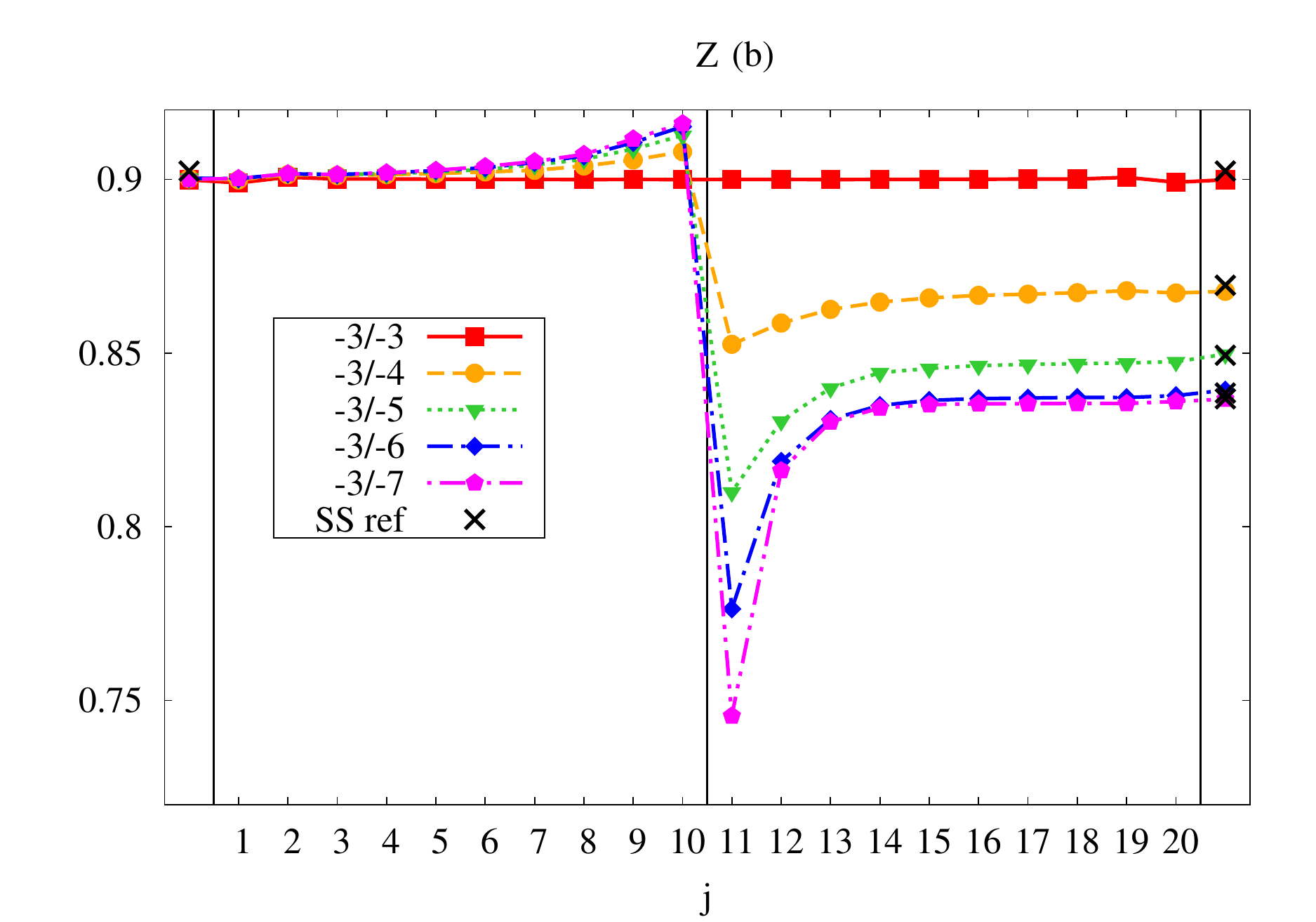}
\includegraphics[scale=0.36]{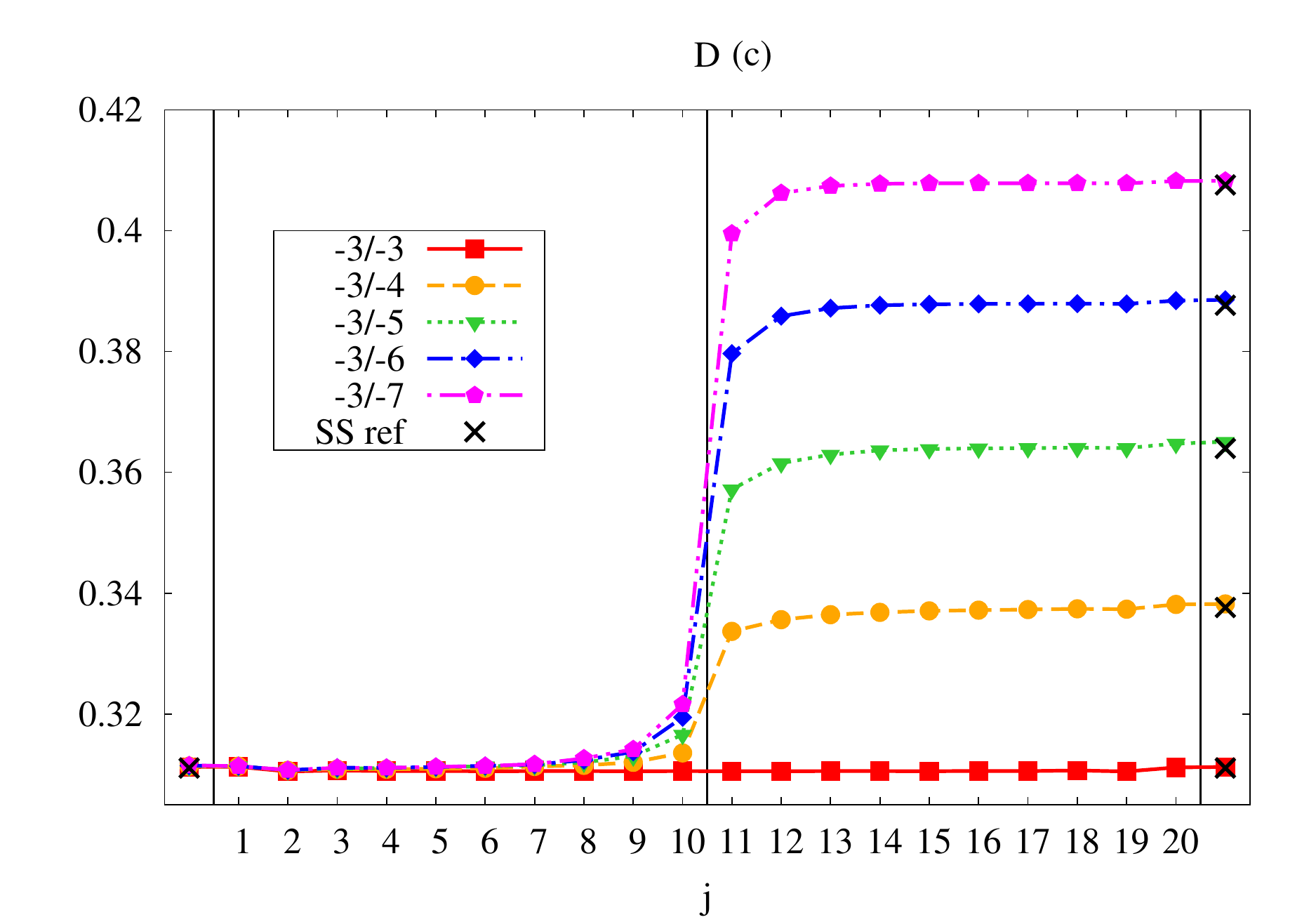}
\includegraphics[scale=0.36]{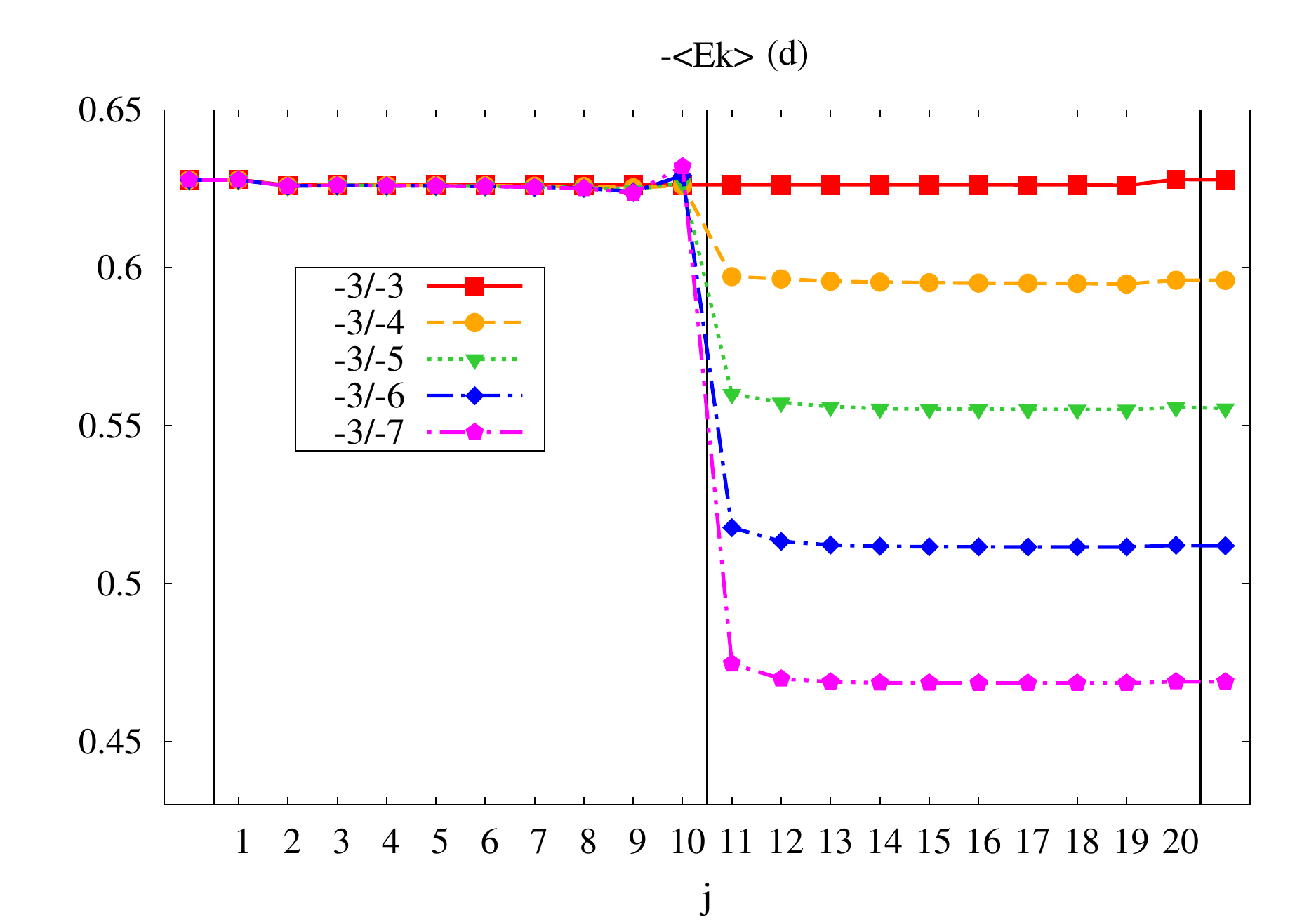}
\caption{$\Delta_{\alpha}$, $z_{\alpha}$, $D_{\alpha}$ and $-\left\langle E_{k}\right\rangle $
  in a $20$-layer thick heterostructure formed by two semi-infinite halves. The left half (index $\leq$ 10) is kept
  at $U/t=-3$, while for the right
  half (index $\geq$ 11) we used different values of the attraction with equal or larger absolute value.
  The points outside the heterostructure are those computed starting from the leads' Green's functions used to compute the embedding potential $\hat{G}_{B}$. The crosses are the results for a bulk DMFT calculation for the cubic lattice.
   \label{fig:3}}
\end{figure}

\begin{figure} 
\includegraphics[scale=0.36]{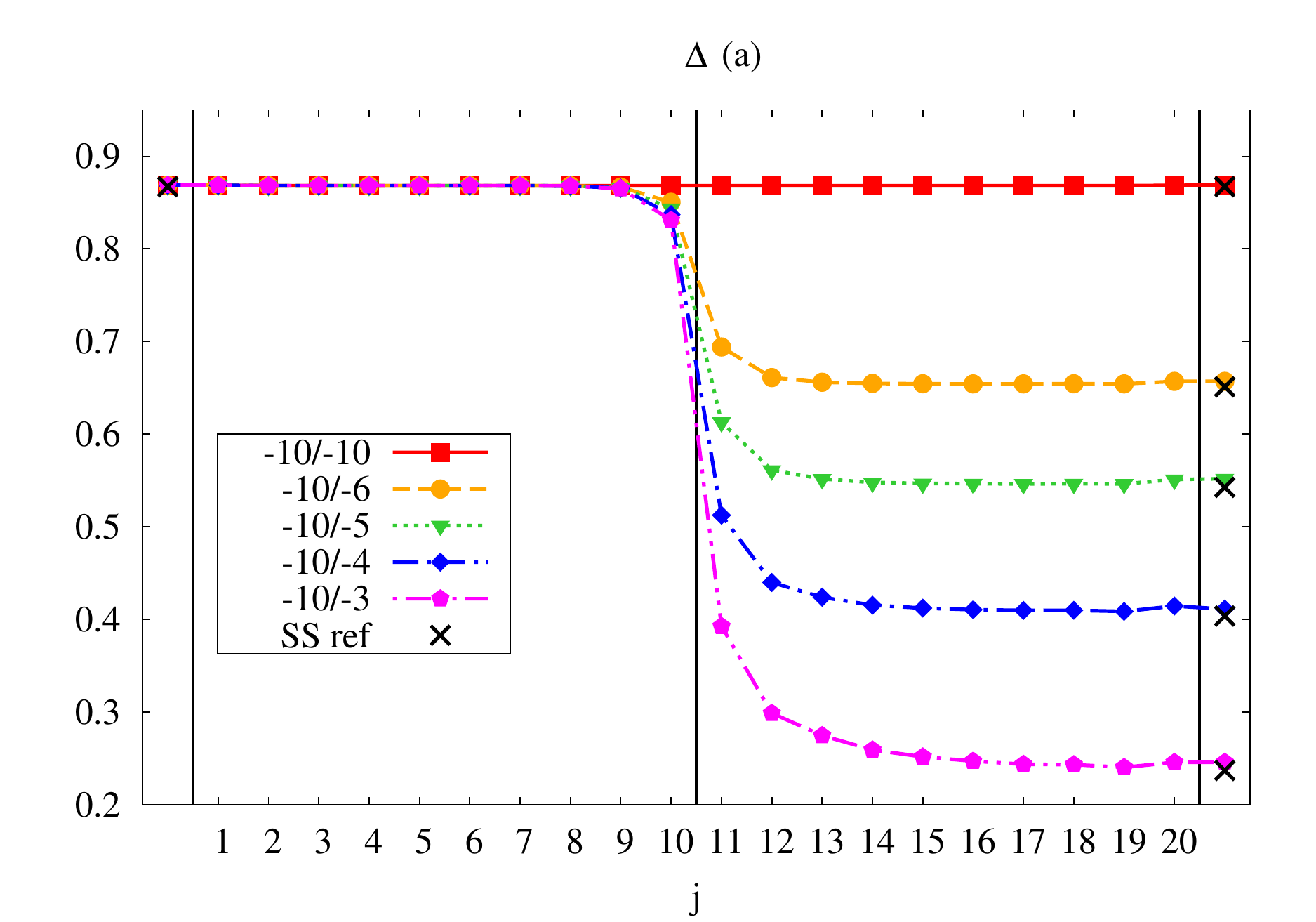}
\includegraphics[scale=0.36]{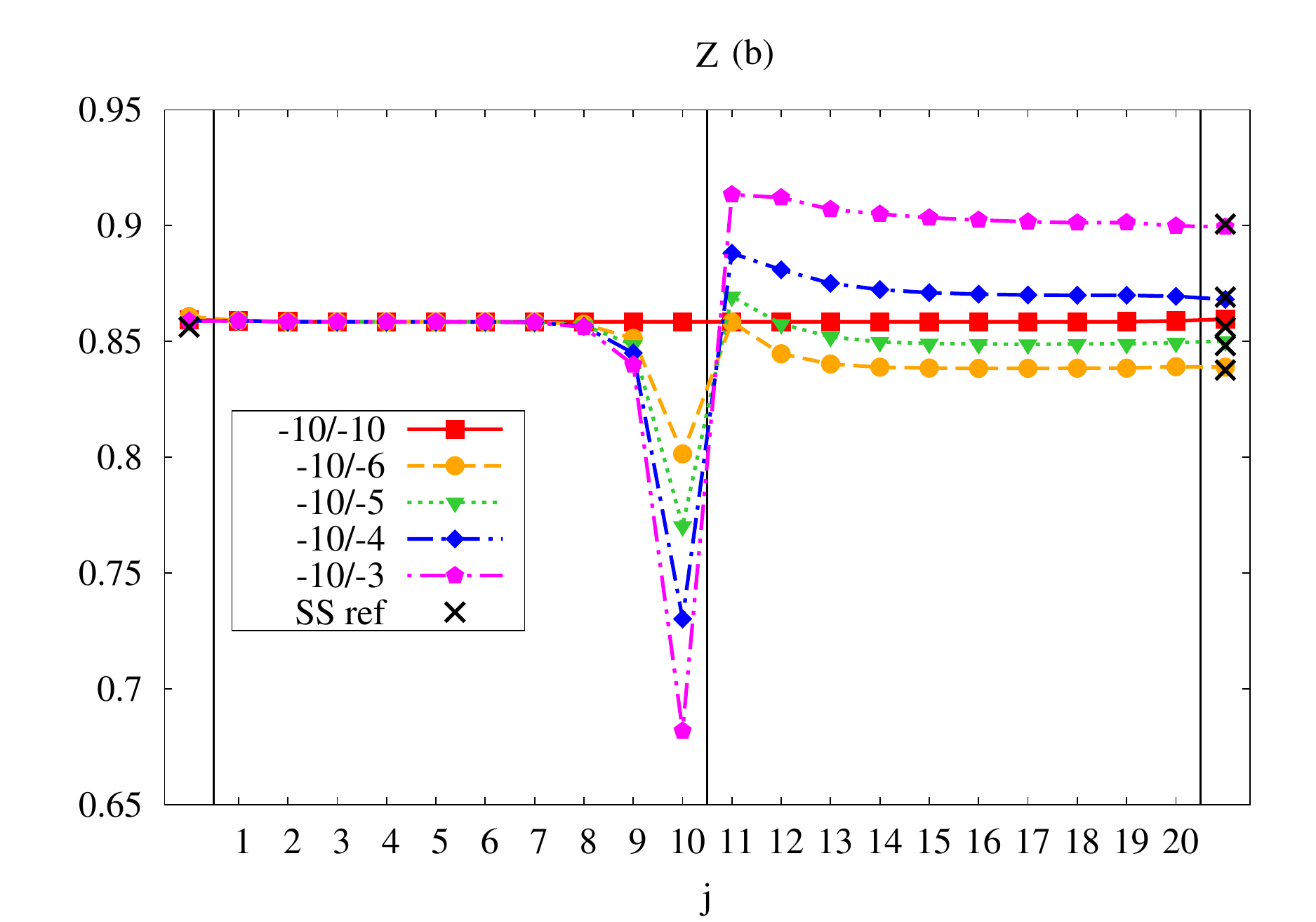}
\includegraphics[scale=0.36]{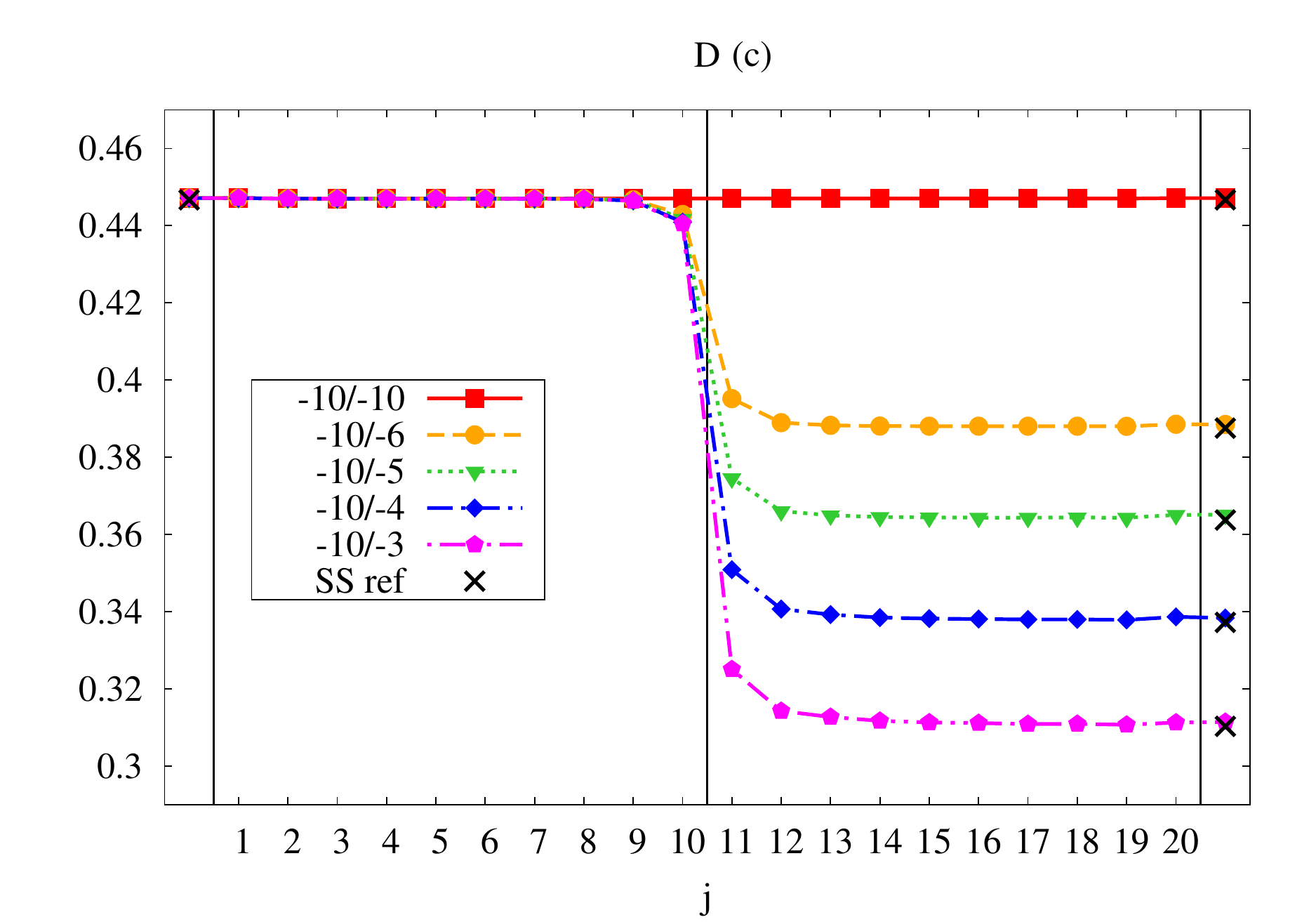}
\includegraphics[scale=0.36]{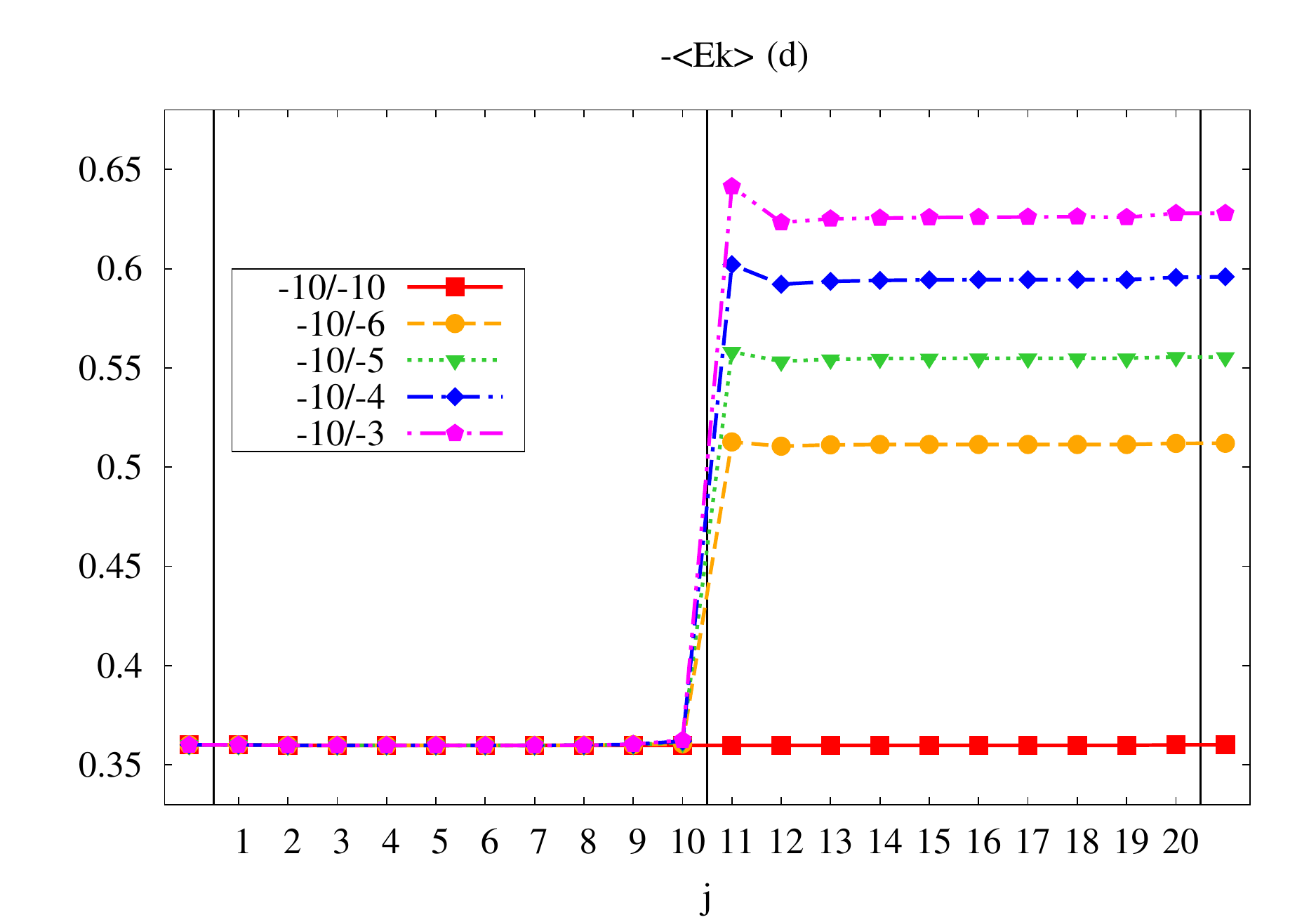}
\caption{$\Delta_{\alpha}$, $z_{\alpha}$, $D_{\alpha}$ and $-\left\langle E_{k}\right\rangle $
in a $20$-layer thick heterostructure. 
 The left half (index $\leq$ 10) is kept
  at $U/t=-10$, while for the right
  half (index $\geq$ 11) we used smaller or equal values of the attraction strength. 
  The points outside the heterostructure are those computed starting from the leads' Green's functions used to compute the embedding potential $\hat{G}_{B}$. The crosses are the results for a bulk DMFT calculation for the cubic lattice. \label{fig:3bis}}
\end{figure}
 
 In this section we present some results using the above defined embedding+feedback procedure for
an attractive Hubbard model. In this work we limit ourselves to paradigmatic situations and we postpone to 
future applications more realistic set-ups corresponding to actual materials and heterostructure. We fix the local density 
to one electron per site on each layer by imposing particle-hole symmetry. This obviously freezes charge redistribution across the 
interface. We chose to start with this situation to single out the intrinsic effects due to the proximity from the effects due to charge 
transfer across the interface, which would obviously affect the results. Interestingly, we find important proximity effect even in this
case.

As a first example we consider the interface between two semi-infinite systems with different values of the attractive interaction, considering ten active layers
for both systems. In Fig. \ref{fig:3}  we present results in which we
fix the interaction at  a relatively small interaction $U/t=-3$  on the left side, while on the right side we tune the interaction from $U/t=-3$ to 
a much larger attraction $U/t =-7.5$. We present layer-resolved  pairing amplitude$\Delta_{\alpha}$,
quasiparticle weight $z_{\alpha}$, double occupancy $D_{\alpha}$ and in-plane kinetic energy $\langle E_{k\alpha}\rangle$
as a function of the layer index $\alpha$. On the right side of the figure the bulk values are shown for reference.

We first observe that also in this case our embedding scheme provides the correct value of every observable in the layers
adjacent to the leads. The evolution across the slab is rather smooth, especially for the order parameter, shown in panel (a), for 
which a significant proximity effect leads to an enhancement of the order parameter on the left side which penetrates for
almost ten layers. Also the right-side is substantially affected by the presence of the weakly-coupled superconductor. 
Interestingly the spatial variation is not strongly dependent on the value of the interaction in the right half.

The double occupancy, which is also related to the potential energy has a similar evolution, but the proximity effects are 
limited to a thinner slice of the slab of around three layers. A similar behavior is shown by the layer kinetic energy, which
is negative and larger in amplitude on the left (weak-coupling side). Interestingly, the presence of the stronger-coupling superconductors
leads to a slight increase of the modulus of the kinetic energy in the first layers of the weak-coupling side.

Finally, the quasiparticle weight, which can be used to measure the coherence of the electronic excitations, is slightly increased in the weak coupling side,
and it decreases in the strong-coupling region, even if all these variations are relatively small. 

In Fig. \ref{fig:3bis} we present an analogous analysis in which the left side has a constant $U/t=-10$ while on the right side the interaction goes from -10 to -3. The qualitative results are similar to the previous even if the proximity effects are reduced because of the stronger coupling on the left side, which leads to a short coherence length and the physics becomes more local. Still, a clear intermediate region in which the physical quantities smoothly connect. 

\subsubsection{Correlated metal/superconductor}
\begin{figure}  
\includegraphics[scale=0.36]{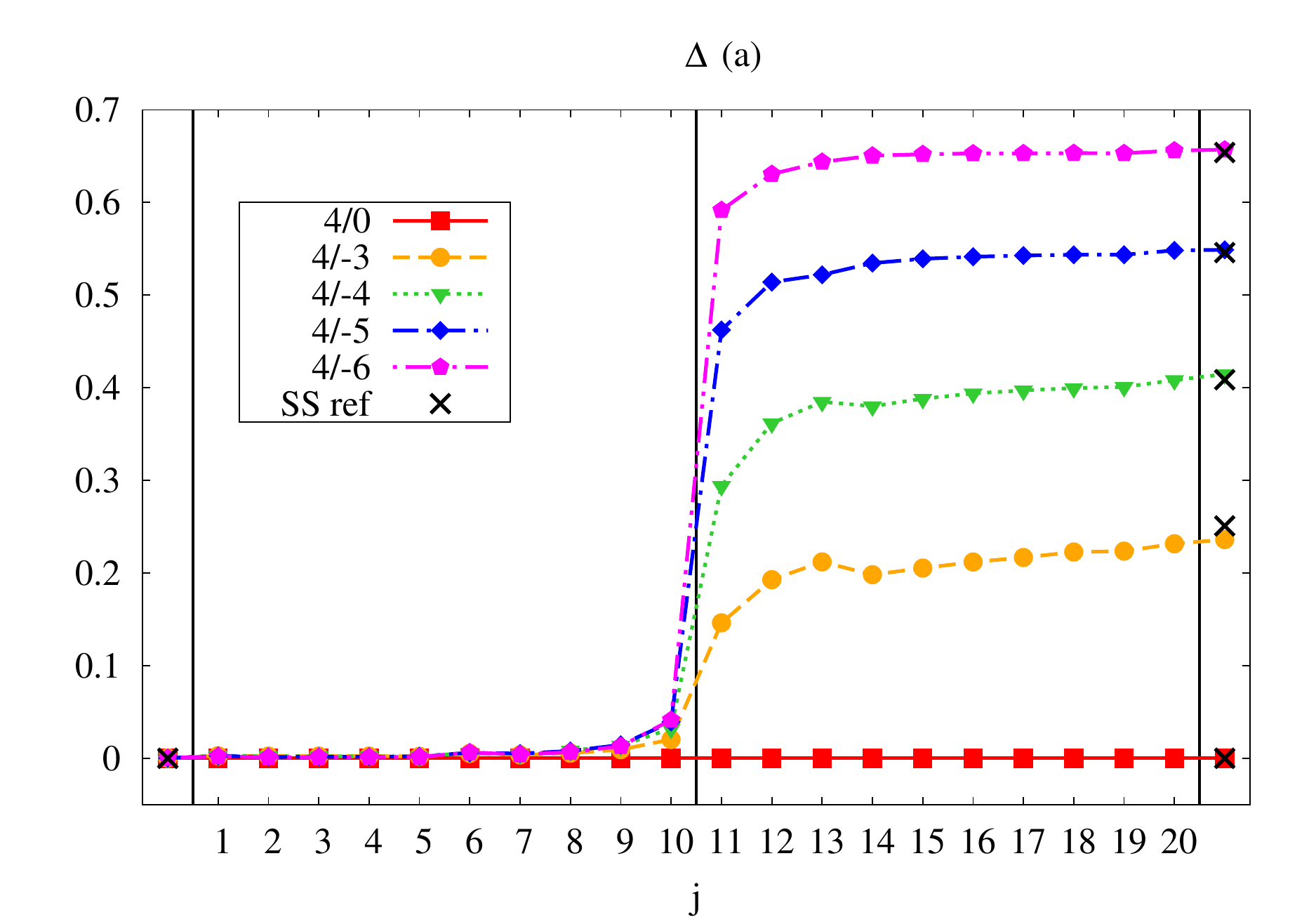}
\includegraphics[scale=0.36]{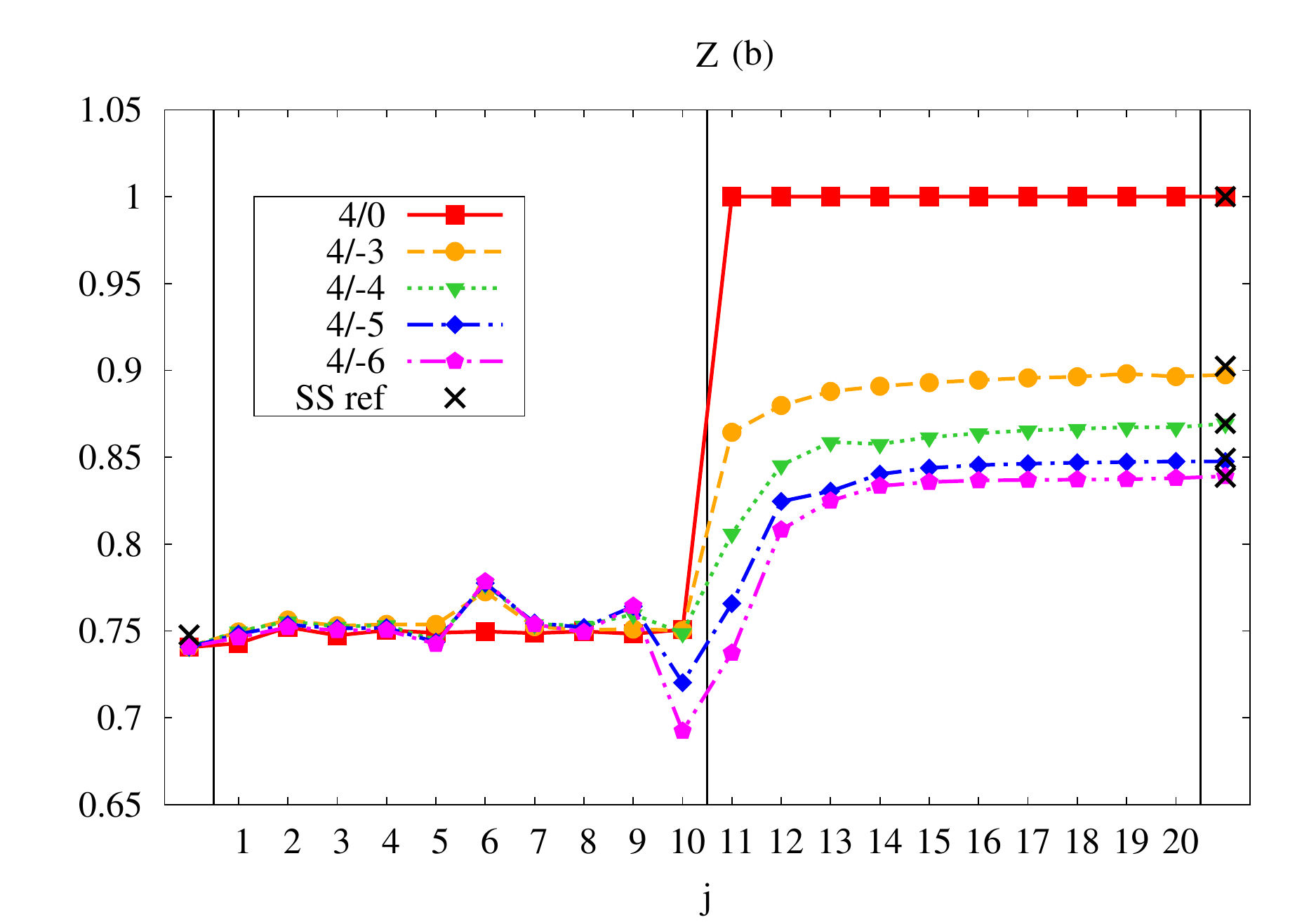}
\includegraphics[scale=0.36]{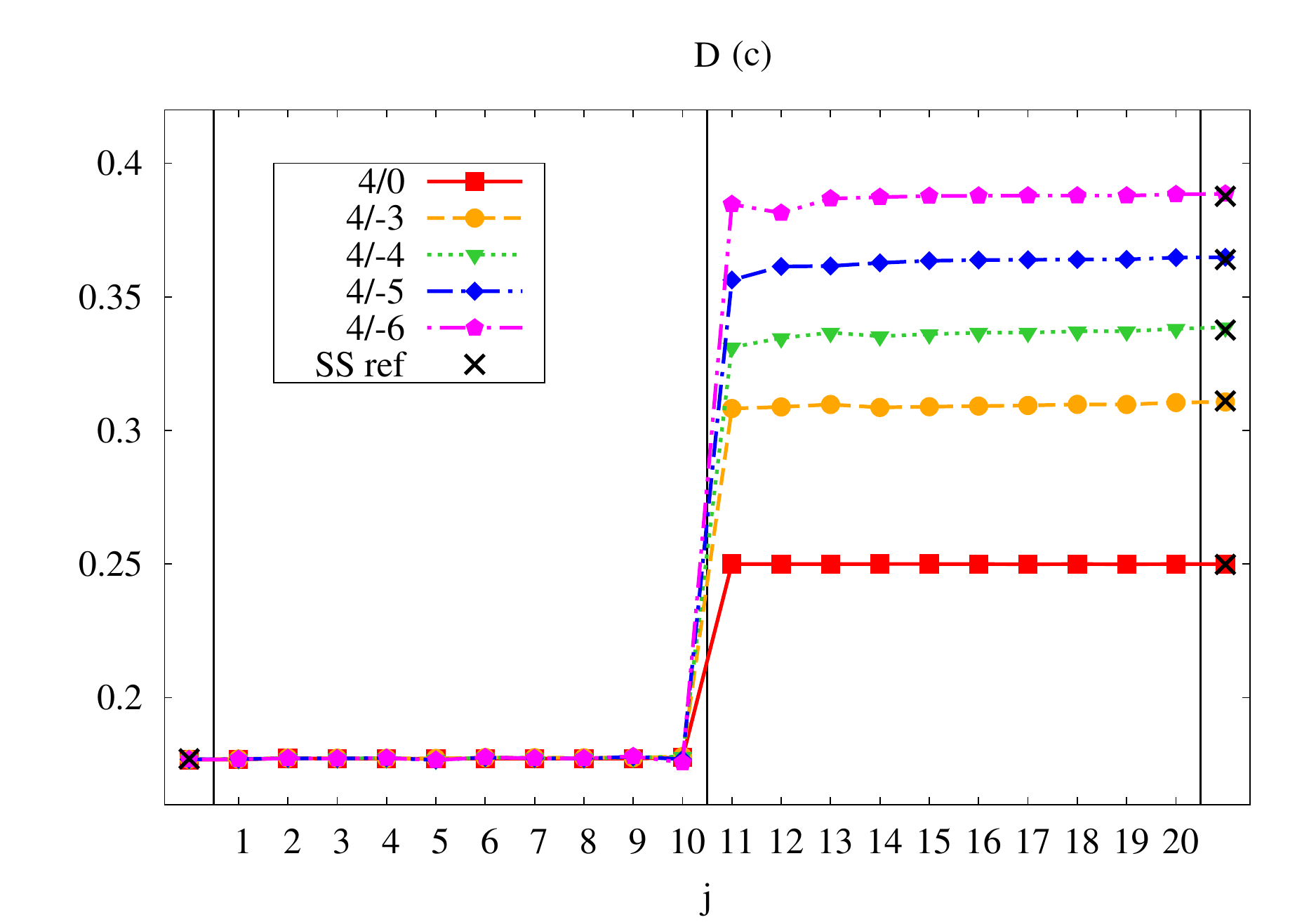}
\includegraphics[scale=0.36]{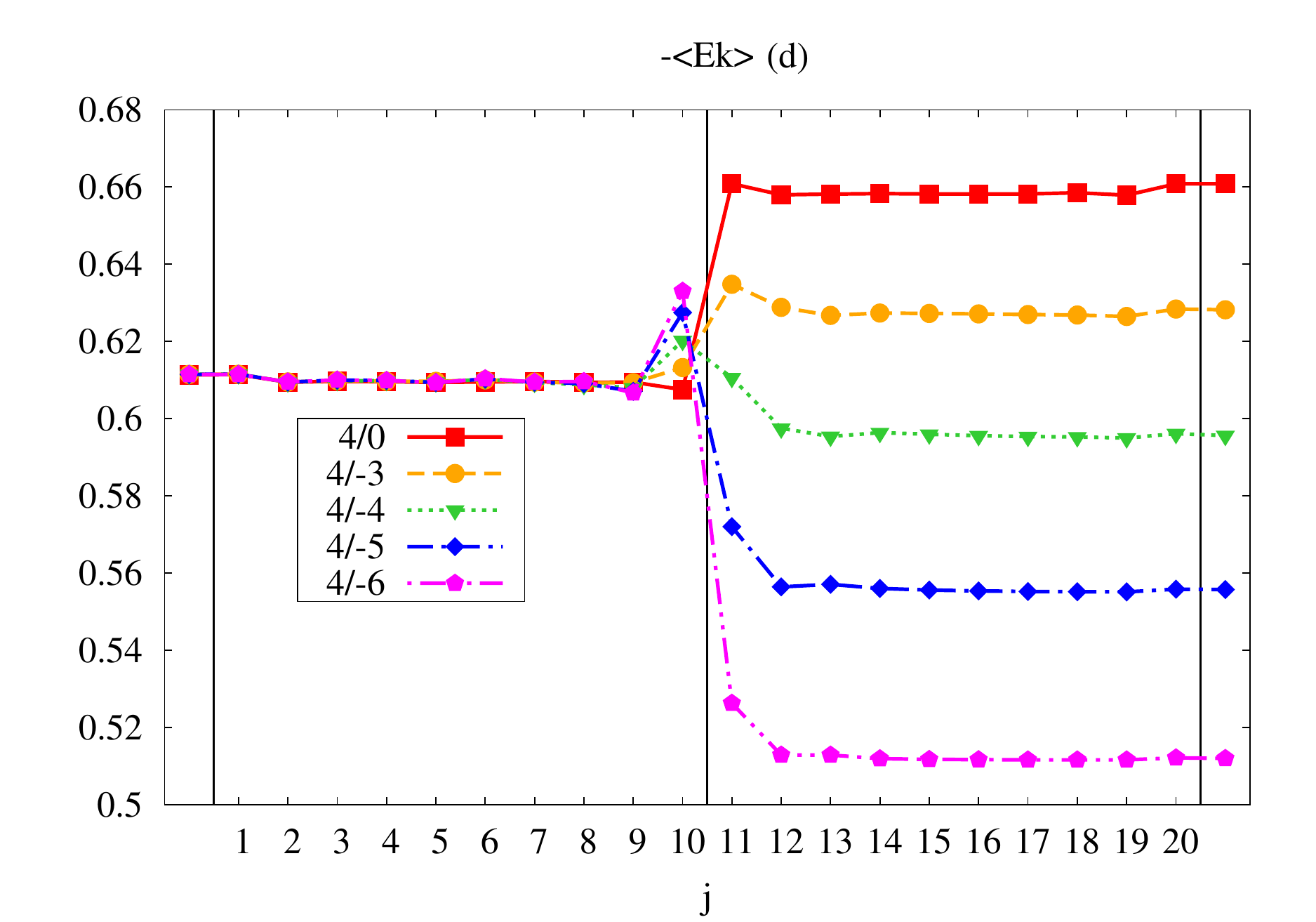}
\caption{$\Delta_{\alpha}$, $z_{\alpha}$, $D_{\alpha}$ and $-\left\langle E_{k}\right\rangle $
in a $20$ layer thick heterostructure. 
Here on the left side we have  a correlate metal with a repulsive $U/t = 4$, while on the right side we tune an attractive interaction. 
  The points outside the heterostructure are those computed starting from the leads' Green's functions used to compute the embedding potential $\hat{G}_{B}$. The crosses are the results for a bulk DMFT calculation for the cubic lattice.\label{fig:4}}
\end{figure}

We now move to a different situation where one of the two halves of the system would not be superconducting by itself.
On the left side we consider a metal with a finite repulsion $U/t = 4$, which would lead to a moderately correlated metal in a bulk system, while on the right we tune the attractive interaction from 0 to $U = -6t$.  The results, plotted in Fig. \ref{fig:4}, show that despite the repulsive interaction 
superconductivity can penetrate for a few layers of the metal, and that important effects are observed 
on the superconducting side. This is a clear qualitative violation of the local-density approximation even in the absence
of charge redistribution across the interface.
The effect on the order parameter is small 
but clearly visible, while the double occupancy is essentially unaffected by the connection between the two semi-infinite systems. The kinetic
energy presents an interesting increase (in modulus) in the first layers of the metallic system, the same region where superconductivity
is able to penetrate in the repulsive metal. 

These results clearly demonstrate that the approach we have devised is able on one hand to reproduce the bulk results when we are sufficiently
far from the interface and on the other hand to display non-trivial and interaction dependent proximity effects, which can lead to important effects
in real systems. The effect is generally stronger for the order parameter.

\section{Conclusion}

In the present work we have introduced an effective extension of the embedding approach
which allows to study heterostructure of interacting systems by means of a small number of 
active layers. Our extension is twofold. For the first time we extend the formalism to the 
superconducting state, and we also introduce a ``feedback" of the slab onto the embedding 
potential describing the rest of the system which reduces the inhomogeneity effects.

This feedback correction has been shown to dramatically reduce the effects of the finiteness of
the slab and to produce essentially exact results for all the relevant layer-resolved observables
observables when treating homogeneous bulk systems within this approximation. 

We have also presented two applications of the method to paradigmatic situations where a 
heterostructure is formed out of two semi-infinite bulks. In particular we consider a superconductor
with different values of the attractive strength connected with either a fixed weak-coupling superconductor
or a metal with intermediate repulsive interactions. We find that in the first case important proximity effects
take place and stronger superconductor increases the superconducting order parameter for around ten 
layers for a wide range of parameters. In the second case superconductivity penetrates in the repulsive system
for around two layers, qualitatively changing the physics of the system. In both cases the strongest effects
are seen on the order parameters, while the kinetic and potential energies remain closer to the results for two 
disconnected systems. It is worth mentioning that, imposing particle-hole symmetry and fixing every layer to be 
half-filled, we freeze the charge redistribution which would naturally enhance the effects we describe. 

\section*{Acknowledgements}
We thank A. Amaricci and G. Giovannetti for useful discussion. This work has been supported by the European Union under FP7 ERC Starting
Grant No. 240524 ``SUPERBAD''.

\bibliography{biblio}

\begin{thebibliography}{37}%
\makeatletter
\providecommand \@ifxundefined [1]{%
 \@ifx{#1\undefined}
}%
\providecommand \@ifnum [1]{%
 \ifnum #1\expandafter \@firstoftwo
 \else \expandafter \@secondoftwo
 \fi
}%
\providecommand \@ifx [1]{%
 \ifx #1\expandafter \@firstoftwo
 \else \expandafter \@secondoftwo
 \fi
}%
\providecommand \natexlab [1]{#1}%
\providecommand \enquote  [1]{``#1''}%
\providecommand \bibnamefont  [1]{#1}%
\providecommand \bibfnamefont [1]{#1}%
\providecommand \citenamefont [1]{#1}%
\providecommand \href@noop [0]{\@secondoftwo}%
\providecommand \href [0]{\begingroup \@sanitize@url \@href}%
\providecommand \@href[1]{\@@startlink{#1}\@@href}%
\providecommand \@@href[1]{\endgroup#1\@@endlink}%
\providecommand \@sanitize@url [0]{\catcode `\\12\catcode `\$12\catcode
  `\&12\catcode `\#12\catcode `\^12\catcode `\_12\catcode `\%12\relax}%
\providecommand \@@startlink[1]{}%
\providecommand \@@endlink[0]{}%
\providecommand \url  [0]{\begingroup\@sanitize@url \@url }%
\providecommand \@url [1]{\endgroup\@href {#1}{\urlprefix }}%
\providecommand \urlprefix  [0]{URL }%
\providecommand \Eprint [0]{\href }%
\@ifxundefined \urlstyle {%
  \providecommand \doi  [0]{\begingroup \@sanitize@url \@doi}%
  \providecommand \@doi [1]{\endgroup \@@startlink {\doibase
  #1}doi:\discretionary {}{}{}#1\@@endlink }%
}{%
  \providecommand \doi  [0]{doi:\discretionary{}{}{}\begingroup
  \urlstyle{rm}\Url }%
}%
\providecommand \doibase [0]{http://dx.doi.org/}%
\providecommand \Doi [0]{\begingroup \@sanitize@url \@Doi }%
\providecommand \@Doi  [1]{\endgroup\@@startlink{\doibase#1}\@@Doi}%
\providecommand \@@Doi [1]{#1\@@endlink}%
\providecommand \selectlanguage [0]{\@gobble}%
\providecommand \bibinfo  [0]{\@secondoftwo}%
\providecommand \bibfield  [0]{\@secondoftwo}%
\providecommand \translation [1]{[#1]}%
\providecommand \BibitemOpen [0]{}%
\providecommand \bibitemStop [0]{}%
\providecommand \bibitemNoStop [0]{.\EOS\space}%
\providecommand \EOS [0]{\spacefactor3000\relax}%
\providecommand \BibitemShut  [1]{\csname bibitem#1\endcsname}%
\bibitem [{\citenamefont {Ohtomo}\ \emph {et~al.}(2010)\citenamefont {Ohtomo},
  \citenamefont {Muller}, \citenamefont {Grazul},\ and\ \citenamefont
  {Hwang}}]{Ohtomo2002}%
  \BibitemOpen
  \bibfield  {author} {\bibinfo {author} {\bibfnamefont {A.}~\bibnamefont
  {Ohtomo}}, \bibinfo {author} {\bibfnamefont {D.~A.}\ \bibnamefont {Muller}},
  \bibinfo {author} {\bibfnamefont {J.~L.}\ \bibnamefont {Grazul}}, \ and\
  \bibinfo {author} {\bibfnamefont {H.}~\bibnamefont {Hwang}},\ }\href
  {http://www.nature.com/ncomms/journal/v1/n7/abs/ncomms1084.html} {\bibfield
  {journal} {\bibinfo  {journal} {Nature},\ }\textbf {\bibinfo {volume} {419}}
  (\bibinfo {year} {2010})}\BibitemShut {NoStop}%
\bibitem [{\citenamefont {Biscaras}\ \emph {et~al.}(2010)\citenamefont
  {Biscaras}, \citenamefont {Bergeal}, \citenamefont {Kushwaha}, \citenamefont
  {Wolf}, \citenamefont {Rastogi}, \citenamefont {Budhani},\ and\ \citenamefont
  {Lesueur}}]{Biscaras2010}%
  \BibitemOpen
  \bibfield  {author} {\bibinfo {author} {\bibfnamefont {J.}~\bibnamefont
  {Biscaras}}, \bibinfo {author} {\bibfnamefont {N.}~\bibnamefont {Bergeal}},
  \bibinfo {author} {\bibfnamefont {A.}~\bibnamefont {Kushwaha}}, \bibinfo
  {author} {\bibfnamefont {T.}~\bibnamefont {Wolf}}, \bibinfo {author}
  {\bibfnamefont {A.}~\bibnamefont {Rastogi}}, \bibinfo {author} {\bibfnamefont
  {R.}~\bibnamefont {Budhani}}, \ and\ \bibinfo {author} {\bibfnamefont
  {J.}~\bibnamefont {Lesueur}},\ }\href
  {http://www.nature.com/ncomms/journal/v1/n7/abs/ncomms1084.html} {\bibfield
  {journal} {\bibinfo  {journal} {Nat. Commun.},\ }\textbf {\bibinfo {volume}
  {1}},\ \bibinfo {pages} {89} (\bibinfo {year} {2010})}\BibitemShut {NoStop}%
\bibitem [{\citenamefont {Biscaras}\ \emph {et~al.}(2012)\citenamefont
  {Biscaras}, \citenamefont {Bergeal}, \citenamefont {Hurand}, \citenamefont
  {Grosset\^ete}, \citenamefont {Rastogi}, \citenamefont {Budhani},
  \citenamefont {LeBoeuf}, \citenamefont {Proust},\ and\ \citenamefont
  {Lesueur}}]{Biscaras2012}%
  \BibitemOpen
  \bibfield  {author} {\bibinfo {author} {\bibfnamefont {J.}~\bibnamefont
  {Biscaras}}, \bibinfo {author} {\bibfnamefont {N.}~\bibnamefont {Bergeal}},
  \bibinfo {author} {\bibfnamefont {S.}~\bibnamefont {Hurand}}, \bibinfo
  {author} {\bibfnamefont {C.}~\bibnamefont {Grosset\^ete}}, \bibinfo {author}
  {\bibfnamefont {A.}~\bibnamefont {Rastogi}}, \bibinfo {author} {\bibfnamefont
  {R.~C.}\ \bibnamefont {Budhani}}, \bibinfo {author} {\bibfnamefont
  {D.}~\bibnamefont {LeBoeuf}}, \bibinfo {author} {\bibfnamefont
  {C.}~\bibnamefont {Proust}}, \ and\ \bibinfo {author} {\bibfnamefont
  {J.}~\bibnamefont {Lesueur}},\ }\Doi {10.1103/PhysRevLett.108.247004}
  {\bibfield  {journal} {\bibinfo  {journal} {Phys. Rev. Lett.},\ }\textbf
  {\bibinfo {volume} {108}},\ \bibinfo {pages} {247004} (\bibinfo {year}
  {2012})}\BibitemShut {NoStop}%
\bibitem [{\citenamefont {Reyren}\ \emph {et~al.}(2007)\citenamefont {Reyren},
  \citenamefont {Thiel}, \citenamefont {Caviglia}, \citenamefont {Kourkoutis},
  \citenamefont {Hammerl}, \citenamefont {Richter}, \citenamefont {Schneider},
  \citenamefont {Kopp}, \citenamefont {Rüetschi}, \citenamefont {Jaccard},
  \citenamefont {Gabay}, \citenamefont {Muller}, \citenamefont {Triscone},\
  and\ \citenamefont {Mannhart}}]{Reyren2007}%
  \BibitemOpen
  \bibfield  {author} {\bibinfo {author} {\bibfnamefont {N.}~\bibnamefont
  {Reyren}}, \bibinfo {author} {\bibfnamefont {S.}~\bibnamefont {Thiel}},
  \bibinfo {author} {\bibfnamefont {A.~D.}\ \bibnamefont {Caviglia}}, \bibinfo
  {author} {\bibfnamefont {L.~F.}\ \bibnamefont {Kourkoutis}}, \bibinfo
  {author} {\bibfnamefont {G.}~\bibnamefont {Hammerl}}, \bibinfo {author}
  {\bibfnamefont {C.}~\bibnamefont {Richter}}, \bibinfo {author} {\bibfnamefont
  {C.~W.}\ \bibnamefont {Schneider}}, \bibinfo {author} {\bibfnamefont
  {T.}~\bibnamefont {Kopp}}, \bibinfo {author} {\bibfnamefont {A.-S.}\
  \bibnamefont {Rüetschi}}, \bibinfo {author} {\bibfnamefont {D.}~\bibnamefont
  {Jaccard}}, \bibinfo {author} {\bibfnamefont {M.}~\bibnamefont {Gabay}},
  \bibinfo {author} {\bibfnamefont {D.~A.}\ \bibnamefont {Muller}}, \bibinfo
  {author} {\bibfnamefont {J.-M.}\ \bibnamefont {Triscone}}, \ and\ \bibinfo
  {author} {\bibfnamefont {J.}~\bibnamefont {Mannhart}},\ }\Doi
  {10.1126/science.1146006} {\bibfield  {journal} {\bibinfo  {journal}
  {Science},\ }\textbf {\bibinfo {volume} {317}},\ \bibinfo {pages} {1196}
  (\bibinfo {year} {2007})}\BibitemShut {NoStop}%
\bibitem [{\citenamefont {Hwang}\ \emph {et~al.}(2012)\citenamefont {Hwang},
  \citenamefont {Iwasa}, \citenamefont {Kawasaki}, \citenamefont {Keimer},
  \citenamefont {Nagaosa},\ and\ \citenamefont {Tokura}}]{Hwang2012}%
  \BibitemOpen
  \bibfield  {author} {\bibinfo {author} {\bibfnamefont {H.}~\bibnamefont
  {Hwang}}, \bibinfo {author} {\bibfnamefont {Y.}~\bibnamefont {Iwasa}},
  \bibinfo {author} {\bibfnamefont {M.}~\bibnamefont {Kawasaki}}, \bibinfo
  {author} {\bibfnamefont {B.}~\bibnamefont {Keimer}}, \bibinfo {author}
  {\bibfnamefont {N.}~\bibnamefont {Nagaosa}}, \ and\ \bibinfo {author}
  {\bibfnamefont {Y.}~\bibnamefont {Tokura}},\ }\href
  {http://www.nature.com/nmat/journal/v11/n2/abs/nmat3223.html} {\bibfield
  {journal} {\bibinfo  {journal} {Nat. Mater.},\ }\textbf {\bibinfo {volume}
  {11}},\ \bibinfo {pages} {103} (\bibinfo {year} {2012})}\BibitemShut
  {NoStop}%
\bibitem [{\citenamefont {Chakhalian}\ \emph {et~al.}(2014)\citenamefont
  {Chakhalian}, \citenamefont {Freeland}, \citenamefont {Millis}, \citenamefont
  {Panagopoulos},\ and\ \citenamefont {Rondinelli}}]{Chakhalian2014}%
  \BibitemOpen
  \bibfield  {author} {\bibinfo {author} {\bibfnamefont {J.}~\bibnamefont
  {Chakhalian}}, \bibinfo {author} {\bibfnamefont {J.~W.}\ \bibnamefont
  {Freeland}}, \bibinfo {author} {\bibfnamefont {A.~J.}\ \bibnamefont
  {Millis}}, \bibinfo {author} {\bibfnamefont {C.}~\bibnamefont
  {Panagopoulos}}, \ and\ \bibinfo {author} {\bibfnamefont {J.~M.}\
  \bibnamefont {Rondinelli}},\ }\Doi {10.1103/RevModPhys.86.1189} {\bibfield
  {journal} {\bibinfo  {journal} {Rev. Mod. Phys.},\ }\textbf {\bibinfo
  {volume} {86}},\ \bibinfo {pages} {1189} (\bibinfo {year}
  {2014})}\BibitemShut {NoStop}%
\bibitem [{\citenamefont {Georges}\ \emph {et~al.}(1996)\citenamefont
  {Georges}, \citenamefont {Kotliar}, \citenamefont {Krauth},\ and\
  \citenamefont {Rozenberg}}]{Georges1996}%
  \BibitemOpen
  \bibfield  {author} {\bibinfo {author} {\bibfnamefont {A.}~\bibnamefont
  {Georges}}, \bibinfo {author} {\bibfnamefont {G.}~\bibnamefont {Kotliar}},
  \bibinfo {author} {\bibfnamefont {W.}~\bibnamefont {Krauth}}, \ and\ \bibinfo
  {author} {\bibfnamefont {M.~J.}\ \bibnamefont {Rozenberg}},\ }\Doi
  {10.1103/RevModPhys.68.13} {\bibfield  {journal} {\bibinfo  {journal} {Rev.
  Mod. Phys.},\ }\textbf {\bibinfo {volume} {68}},\ \bibinfo {pages} {13}
  (\bibinfo {year} {1996})}\BibitemShut {NoStop}%
\bibitem [{\citenamefont {Freericks}\ \emph {et~al.}(1993)\citenamefont
  {Freericks}, \citenamefont {Jarrell},\ and\ \citenamefont
  {Scalapino}}]{Freericks1993}%
  \BibitemOpen
  \bibfield  {author} {\bibinfo {author} {\bibfnamefont {J.~K.}\ \bibnamefont
  {Freericks}}, \bibinfo {author} {\bibfnamefont {M.}~\bibnamefont {Jarrell}},
  \ and\ \bibinfo {author} {\bibfnamefont {D.~J.}\ \bibnamefont {Scalapino}},\
  }\Doi {10.1103/PhysRevB.48.6302} {\bibfield  {journal} {\bibinfo  {journal}
  {Phys. Rev. B},\ }\textbf {\bibinfo {volume} {48}},\ \bibinfo {pages} {6302}
  (\bibinfo {year} {1993})}\BibitemShut {NoStop}%
\bibitem [{\citenamefont {Millis}\ \emph {et~al.}(1996)\citenamefont {Millis},
  \citenamefont {Mueller},\ and\ \citenamefont {Shraiman}}]{Millis1996}%
  \BibitemOpen
  \bibfield  {author} {\bibinfo {author} {\bibfnamefont {A.~J.}\ \bibnamefont
  {Millis}}, \bibinfo {author} {\bibfnamefont {R.}~\bibnamefont {Mueller}}, \
  and\ \bibinfo {author} {\bibfnamefont {B.~I.}\ \bibnamefont {Shraiman}},\
  }\Doi {10.1103/PhysRevB.54.5389} {\bibfield  {journal} {\bibinfo  {journal}
  {Phys. Rev. B},\ }\textbf {\bibinfo {volume} {54}},\ \bibinfo {pages} {5389}
  (\bibinfo {year} {1996})}\BibitemShut {NoStop}%
\bibitem [{\citenamefont {Meyer}\ \emph {et~al.}(2002)\citenamefont {Meyer},
  \citenamefont {Hewson},\ and\ \citenamefont {Bulla}}]{Meyer2002}%
  \BibitemOpen
  \bibfield  {author} {\bibinfo {author} {\bibfnamefont {D.}~\bibnamefont
  {Meyer}}, \bibinfo {author} {\bibfnamefont {A.~C.}\ \bibnamefont {Hewson}}, \
  and\ \bibinfo {author} {\bibfnamefont {R.}~\bibnamefont {Bulla}},\ }\Doi
  {10.1103/PhysRevLett.89.196401} {\bibfield  {journal} {\bibinfo  {journal}
  {Phys. Rev. Lett.},\ }\textbf {\bibinfo {volume} {89}},\ \bibinfo {pages}
  {196401} (\bibinfo {year} {2002})}\BibitemShut {NoStop}%
\bibitem [{\citenamefont {Capone}\ and\ \citenamefont
  {Ciuchi}(2003)}]{Capone2003}%
  \BibitemOpen
  \bibfield  {author} {\bibinfo {author} {\bibfnamefont {M.}~\bibnamefont
  {Capone}}\ and\ \bibinfo {author} {\bibfnamefont {S.}~\bibnamefont
  {Ciuchi}},\ }\Doi {10.1103/PhysRevLett.91.186405} {\bibfield  {journal}
  {\bibinfo  {journal} {Phys. Rev. Lett.},\ }\textbf {\bibinfo {volume} {91}},\
  \bibinfo {pages} {186405} (\bibinfo {year} {2003})}\BibitemShut {NoStop}%
\bibitem [{\citenamefont {Freericks}\ and\ \citenamefont
  {Jarrell}(1995)}]{Freericks1995}%
  \BibitemOpen
  \bibfield  {author} {\bibinfo {author} {\bibfnamefont {J.~K.}\ \bibnamefont
  {Freericks}}\ and\ \bibinfo {author} {\bibfnamefont {M.}~\bibnamefont
  {Jarrell}},\ }\Doi {10.1103/PhysRevLett.75.2570} {\bibfield  {journal}
  {\bibinfo  {journal} {Phys. Rev. Lett.},\ }\textbf {\bibinfo {volume} {75}},\
  \bibinfo {pages} {2570} (\bibinfo {year} {1995})}\BibitemShut {NoStop}%
\bibitem [{\citenamefont {Deppeler}\ and\ \citenamefont
  {Millis}(2002)}]{Deppeler2002}%
  \BibitemOpen
  \bibfield  {author} {\bibinfo {author} {\bibfnamefont {A.}~\bibnamefont
  {Deppeler}}\ and\ \bibinfo {author} {\bibfnamefont {A.~J.}\ \bibnamefont
  {Millis}},\ }\Doi {10.1103/PhysRevB.65.100301} {\bibfield  {journal}
  {\bibinfo  {journal} {Phys. Rev. B},\ }\textbf {\bibinfo {volume} {65}},\
  \bibinfo {pages} {100301} (\bibinfo {year} {2002})}\BibitemShut {NoStop}%
\bibitem [{\citenamefont {Koller}\ \emph {et~al.}(2004)\citenamefont {Koller},
  \citenamefont {Meyer},\ and\ \citenamefont {Hewson}}]{Koller2004}%
  \BibitemOpen
  \bibfield  {author} {\bibinfo {author} {\bibfnamefont {W.}~\bibnamefont
  {Koller}}, \bibinfo {author} {\bibfnamefont {D.}~\bibnamefont {Meyer}}, \
  and\ \bibinfo {author} {\bibfnamefont {A.~C.}\ \bibnamefont {Hewson}},\ }\Doi
  {10.1103/PhysRevB.70.155103} {\bibfield  {journal} {\bibinfo  {journal}
  {Phys. Rev. B},\ }\textbf {\bibinfo {volume} {70}},\ \bibinfo {pages}
  {155103} (\bibinfo {year} {2004})}\BibitemShut {NoStop}%
\bibitem [{\citenamefont {Sangiovanni}\ \emph {et~al.}(2005)\citenamefont
  {Sangiovanni}, \citenamefont {Capone}, \citenamefont {Castellani},\ and\
  \citenamefont {Grilli}}]{Sangiovanni2005}%
  \BibitemOpen
  \bibfield  {author} {\bibinfo {author} {\bibfnamefont {G.}~\bibnamefont
  {Sangiovanni}}, \bibinfo {author} {\bibfnamefont {M.}~\bibnamefont {Capone}},
  \bibinfo {author} {\bibfnamefont {C.}~\bibnamefont {Castellani}}, \ and\
  \bibinfo {author} {\bibfnamefont {M.}~\bibnamefont {Grilli}},\ }\Doi
  {10.1103/PhysRevLett.94.026401} {\bibfield  {journal} {\bibinfo  {journal}
  {Phys. Rev. Lett.},\ }\textbf {\bibinfo {volume} {94}},\ \bibinfo {pages}
  {026401} (\bibinfo {year} {2005})}\BibitemShut {NoStop}%
\bibitem [{\citenamefont {Werner}\ and\ \citenamefont
  {Millis}(2007)}]{Werner2007}%
  \BibitemOpen
  \bibfield  {author} {\bibinfo {author} {\bibfnamefont {P.}~\bibnamefont
  {Werner}}\ and\ \bibinfo {author} {\bibfnamefont {A.~J.}\ \bibnamefont
  {Millis}},\ }\Doi {10.1103/PhysRevLett.99.146404} {\bibfield  {journal}
  {\bibinfo  {journal} {Phys. Rev. Lett.},\ }\textbf {\bibinfo {volume} {99}},\
  \bibinfo {pages} {146404} (\bibinfo {year} {2007})}\BibitemShut {NoStop}%
\bibitem [{\citenamefont {Sangiovanni}\ \emph {et~al.}(2006)\citenamefont
  {Sangiovanni}, \citenamefont {Capone},\ and\ \citenamefont
  {Castellani}}]{Sangiovanni2006}%
  \BibitemOpen
  \bibfield  {author} {\bibinfo {author} {\bibfnamefont {G.}~\bibnamefont
  {Sangiovanni}}, \bibinfo {author} {\bibfnamefont {M.}~\bibnamefont {Capone}},
  \ and\ \bibinfo {author} {\bibfnamefont {C.}~\bibnamefont {Castellani}},\
  }\Doi {10.1103/PhysRevB.73.165123} {\bibfield  {journal} {\bibinfo  {journal}
  {Phys. Rev. B},\ }\textbf {\bibinfo {volume} {73}},\ \bibinfo {pages}
  {165123} (\bibinfo {year} {2006})}\BibitemShut {NoStop}%
\bibitem [{\citenamefont {Capone}\ \emph {et~al.}(2010)\citenamefont {Capone},
  \citenamefont {Castellani},\ and\ \citenamefont {Grilli}}]{Capone2010}%
  \BibitemOpen
  \bibfield  {author} {\bibinfo {author} {\bibfnamefont {M.}~\bibnamefont
  {Capone}}, \bibinfo {author} {\bibfnamefont {C.}~\bibnamefont {Castellani}},
  \ and\ \bibinfo {author} {\bibfnamefont {M.}~\bibnamefont {Grilli}},\ }\Doi
  {10.1155/2010/9208601} {\bibfield  {journal} {\bibinfo  {journal} {Advances
  in Condensed Matter Physics},\ }\textbf {\bibinfo {volume} {2010}},\ \bibinfo
  {pages} {9208601} (\bibinfo {year} {2010})}\BibitemShut {NoStop}%
\bibitem [{\citenamefont {Giovannetti}\ \emph {et~al.}(2014)\citenamefont
  {Giovannetti}, \citenamefont {Casula}, \citenamefont {Werner}, \citenamefont
  {Mauri},\ and\ \citenamefont {Capone}}]{Giovannetti2014}%
  \BibitemOpen
  \bibfield  {author} {\bibinfo {author} {\bibfnamefont {G.}~\bibnamefont
  {Giovannetti}}, \bibinfo {author} {\bibfnamefont {M.}~\bibnamefont {Casula}},
  \bibinfo {author} {\bibfnamefont {P.}~\bibnamefont {Werner}}, \bibinfo
  {author} {\bibfnamefont {F.}~\bibnamefont {Mauri}}, \ and\ \bibinfo {author}
  {\bibfnamefont {M.}~\bibnamefont {Capone}},\ }\Doi
  {10.1103/PhysRevB.90.115435} {\bibfield  {journal} {\bibinfo  {journal}
  {Phys. Rev. B},\ }\textbf {\bibinfo {volume} {90}},\ \bibinfo {pages}
  {115435} (\bibinfo {year} {2014})}\BibitemShut {NoStop}%
\bibitem [{\citenamefont {Nomura}\ \emph {et~al.}(2015)\citenamefont {Nomura},
  \citenamefont {Sakai}, \citenamefont {Capone},\ and\ \citenamefont
  {Arita}}]{Nomura2015}%
  \BibitemOpen
  \bibfield  {author} {\bibinfo {author} {\bibfnamefont {Y.}~\bibnamefont
  {Nomura}}, \bibinfo {author} {\bibfnamefont {S.}~\bibnamefont {Sakai}},
  \bibinfo {author} {\bibfnamefont {M.}~\bibnamefont {Capone}}, \ and\ \bibinfo
  {author} {\bibfnamefont {R.}~\bibnamefont {Arita}},\ }\href@noop {}
  {\bibfield  {journal} {\bibinfo  {journal} {Science Advances},\ }\textbf
  {\bibinfo {volume} {1}} (\bibinfo {year} {2015})}\BibitemShut {NoStop}%
\bibitem [{\citenamefont {Keller}\ \emph {et~al.}(2001)\citenamefont {Keller},
  \citenamefont {Metzner},\ and\ \citenamefont {Schollw\"ock}}]{Keller2001}%
  \BibitemOpen
  \bibfield  {author} {\bibinfo {author} {\bibfnamefont {M.}~\bibnamefont
  {Keller}}, \bibinfo {author} {\bibfnamefont {W.}~\bibnamefont {Metzner}}, \
  and\ \bibinfo {author} {\bibfnamefont {U.}~\bibnamefont {Schollw\"ock}},\
  }\Doi {10.1103/PhysRevLett.86.4612} {\bibfield  {journal} {\bibinfo
  {journal} {Phys. Rev. Lett.},\ }\textbf {\bibinfo {volume} {86}},\ \bibinfo
  {pages} {4612} (\bibinfo {year} {2001})}\BibitemShut {NoStop}%
\bibitem [{\citenamefont {Capone}\ \emph {et~al.}(2002)\citenamefont {Capone},
  \citenamefont {Castellani},\ and\ \citenamefont {Grilli}}]{Capone2002}%
  \BibitemOpen
  \bibfield  {author} {\bibinfo {author} {\bibfnamefont {M.}~\bibnamefont
  {Capone}}, \bibinfo {author} {\bibfnamefont {C.}~\bibnamefont {Castellani}},
  \ and\ \bibinfo {author} {\bibfnamefont {M.}~\bibnamefont {Grilli}},\ }\Doi
  {10.1103/PhysRevLett.88.126403} {\bibfield  {journal} {\bibinfo  {journal}
  {Phys. Rev. Lett.},\ }\textbf {\bibinfo {volume} {88}},\ \bibinfo {pages}
  {126403} (\bibinfo {year} {2002})}\BibitemShut {NoStop}%
\bibitem [{\citenamefont {Toschi}\ \emph
  {et~al.}(2005){\natexlab{a}}\citenamefont {Toschi}, \citenamefont {Capone},\
  and\ \citenamefont {Castellani}}]{Toschi2005a}%
  \BibitemOpen
  \bibfield  {author} {\bibinfo {author} {\bibfnamefont {A.}~\bibnamefont
  {Toschi}}, \bibinfo {author} {\bibfnamefont {M.}~\bibnamefont {Capone}}, \
  and\ \bibinfo {author} {\bibfnamefont {C.}~\bibnamefont {Castellani}},\ }\Doi
  {10.1103/PhysRevB.72.235118} {\bibfield  {journal} {\bibinfo  {journal}
  {Phys. Rev. B},\ }\textbf {\bibinfo {volume} {72}},\ \bibinfo {pages}
  {235118} (\bibinfo {year} {2005}{\natexlab{a}})}\BibitemShut {NoStop}%
\bibitem [{\citenamefont {Toschi}\ \emph
  {et~al.}(2005){\natexlab{b}}\citenamefont {Toschi}, \citenamefont {Barone},
  \citenamefont {Capone},\ and\ \citenamefont {Castellani}}]{Toschi2005b}%
  \BibitemOpen
  \bibfield  {author} {\bibinfo {author} {\bibfnamefont {A.}~\bibnamefont
  {Toschi}}, \bibinfo {author} {\bibfnamefont {P.}~\bibnamefont {Barone}},
  \bibinfo {author} {\bibfnamefont {M.}~\bibnamefont {Capone}}, \ and\ \bibinfo
  {author} {\bibfnamefont {C.}~\bibnamefont {Castellani}},\ }\href
  {http://stacks.iop.org/1367-2630/7/i=1/a=007} {\bibfield  {journal} {\bibinfo
   {journal} {New Journal of Physics},\ }\textbf {\bibinfo {volume} {7}},\
  \bibinfo {pages} {7} (\bibinfo {year} {2005}{\natexlab{b}})}\BibitemShut
  {NoStop}%
\bibitem [{\citenamefont {Garg}\ \emph {et~al.}(2005)\citenamefont {Garg},
  \citenamefont {Krishnamurthy},\ and\ \citenamefont {Randeria}}]{Garg2005}%
  \BibitemOpen
  \bibfield  {author} {\bibinfo {author} {\bibfnamefont {A.}~\bibnamefont
  {Garg}}, \bibinfo {author} {\bibfnamefont {H.~R.}\ \bibnamefont
  {Krishnamurthy}}, \ and\ \bibinfo {author} {\bibfnamefont {M.}~\bibnamefont
  {Randeria}},\ }\Doi {10.1103/PhysRevB.72.024517} {\bibfield  {journal}
  {\bibinfo  {journal} {Phys. Rev. B},\ }\textbf {\bibinfo {volume} {72}},\
  \bibinfo {pages} {024517} (\bibinfo {year} {2005})}\BibitemShut {NoStop}%
\bibitem [{\citenamefont {Bauer}\ and\ \citenamefont
  {Hewson}(2009)}]{Bauer2009a}%
  \BibitemOpen
  \bibfield  {author} {\bibinfo {author} {\bibfnamefont {J.}~\bibnamefont
  {Bauer}}\ and\ \bibinfo {author} {\bibfnamefont {A.~C.}\ \bibnamefont
  {Hewson}},\ }\href {http://stacks.iop.org/0295-5075/85/i=2/a=27001}
  {\bibfield  {journal} {\bibinfo  {journal} {EPL (Europhysics Letters)},\
  }\textbf {\bibinfo {volume} {85}},\ \bibinfo {pages} {27001} (\bibinfo {year}
  {2009})}\BibitemShut {NoStop}%
\bibitem [{\citenamefont {Bauer}\ \emph {et~al.}(2009)\citenamefont {Bauer},
  \citenamefont {Hewson},\ and\ \citenamefont {Dupuis}}]{Bauer2009b}%
  \BibitemOpen
  \bibfield  {author} {\bibinfo {author} {\bibfnamefont {J.}~\bibnamefont
  {Bauer}}, \bibinfo {author} {\bibfnamefont {A.~C.}\ \bibnamefont {Hewson}}, \
  and\ \bibinfo {author} {\bibfnamefont {N.}~\bibnamefont {Dupuis}},\ }\Doi
  {10.1103/PhysRevB.79.214518} {\bibfield  {journal} {\bibinfo  {journal}
  {Phys. Rev. B},\ }\textbf {\bibinfo {volume} {79}},\ \bibinfo {pages}
  {214518} (\bibinfo {year} {2009})}\BibitemShut {NoStop}%
\bibitem [{\citenamefont {Koga}\ and\ \citenamefont {Werner}(2011)}]{Koga2011}%
  \BibitemOpen
  \bibfield  {author} {\bibinfo {author} {\bibfnamefont {A.}~\bibnamefont
  {Koga}}\ and\ \bibinfo {author} {\bibfnamefont {P.}~\bibnamefont {Werner}},\
  }\Doi {10.1103/PhysRevA.84.023638} {\bibfield  {journal} {\bibinfo  {journal}
  {Phys. Rev. A},\ }\textbf {\bibinfo {volume} {84}},\ \bibinfo {pages}
  {023638} (\bibinfo {year} {2011})}\BibitemShut {NoStop}%
\bibitem [{\citenamefont {Potthoff}\ and\ \citenamefont
  {Nolting}(1999){\natexlab{a}}}]{Potthoff1999a}%
  \BibitemOpen
  \bibfield  {author} {\bibinfo {author} {\bibfnamefont {M.}~\bibnamefont
  {Potthoff}}\ and\ \bibinfo {author} {\bibfnamefont {W.}~\bibnamefont
  {Nolting}},\ }\Doi {10.1103/PhysRevB.59.2549} {\bibfield  {journal} {\bibinfo
   {journal} {Phys. Rev. B},\ }\textbf {\bibinfo {volume} {59}},\ \bibinfo
  {pages} {2549} (\bibinfo {year} {1999}{\natexlab{a}})}\BibitemShut {NoStop}%
\bibitem [{\citenamefont {Potthoff}\ and\ \citenamefont
  {Nolting}(1999){\natexlab{b}}}]{Potthoff1999b}%
  \BibitemOpen
  \bibfield  {author} {\bibinfo {author} {\bibfnamefont {M.}~\bibnamefont
  {Potthoff}}\ and\ \bibinfo {author} {\bibfnamefont {W.}~\bibnamefont
  {Nolting}},\ }\Doi {10.1103/PhysRevB.60.7834} {\bibfield  {journal} {\bibinfo
   {journal} {Phys. Rev. B},\ }\textbf {\bibinfo {volume} {60}},\ \bibinfo
  {pages} {7834} (\bibinfo {year} {1999}{\natexlab{b}})}\BibitemShut {NoStop}%
\bibitem [{\citenamefont {Okamoto}\ and\ \citenamefont
  {Millis}(2004){\natexlab{a}}}]{Okamoto2004a}%
  \BibitemOpen
  \bibfield  {author} {\bibinfo {author} {\bibfnamefont {S.}~\bibnamefont
  {Okamoto}}\ and\ \bibinfo {author} {\bibfnamefont {A.~J.}\ \bibnamefont
  {Millis}},\ }\Doi {10.1103/PhysRevB.70.241104} {\bibfield  {journal}
  {\bibinfo  {journal} {Phys. Rev. B},\ }\textbf {\bibinfo {volume} {70}},\
  \bibinfo {pages} {241104} (\bibinfo {year} {2004}{\natexlab{a}})}\BibitemShut
  {NoStop}%
\bibitem [{\citenamefont {Okamoto}\ and\ \citenamefont
  {Millis}(2004){\natexlab{b}}}]{Okamoto2004b}%
  \BibitemOpen
  \bibfield  {author} {\bibinfo {author} {\bibfnamefont {S.}~\bibnamefont
  {Okamoto}}\ and\ \bibinfo {author} {\bibfnamefont {A.~J.}\ \bibnamefont
  {Millis}},\ }\Doi {10.1103/PhysRevB.70.075101} {\bibfield  {journal}
  {\bibinfo  {journal} {Phys. Rev. B},\ }\textbf {\bibinfo {volume} {70}},\
  \bibinfo {pages} {075101} (\bibinfo {year} {2004}{\natexlab{b}})}\BibitemShut
  {NoStop}%
\bibitem [{\citenamefont {Kancharla}\ and\ \citenamefont
  {Dagotto}(2006)}]{Kancharla2006}%
  \BibitemOpen
  \bibfield  {author} {\bibinfo {author} {\bibfnamefont {S.~S.}\ \bibnamefont
  {Kancharla}}\ and\ \bibinfo {author} {\bibfnamefont {E.}~\bibnamefont
  {Dagotto}},\ }\Doi {10.1103/PhysRevB.74.195427} {\bibfield  {journal}
  {\bibinfo  {journal} {Phys. Rev. B},\ }\textbf {\bibinfo {volume} {74}},\
  \bibinfo {pages} {195427} (\bibinfo {year} {2006})}\BibitemShut {NoStop}%
\bibitem [{\citenamefont {Ishida}\ and\ \citenamefont
  {Liebsch}(2009)}]{Ishida2009}%
  \BibitemOpen
  \bibfield  {author} {\bibinfo {author} {\bibfnamefont {H.}~\bibnamefont
  {Ishida}}\ and\ \bibinfo {author} {\bibfnamefont {A.}~\bibnamefont
  {Liebsch}},\ }\Doi {10.1103/PhysRevB.79.045130} {\bibfield  {journal}
  {\bibinfo  {journal} {Phys. Rev. B},\ }\textbf {\bibinfo {volume} {79}},\
  \bibinfo {pages} {045130} (\bibinfo {year} {2009})}\BibitemShut {NoStop}%
\bibitem [{\citenamefont {Ishida}\ and\ \citenamefont
  {Liebsch}(2010)}]{Ishida2010}%
  \BibitemOpen
  \bibfield  {author} {\bibinfo {author} {\bibfnamefont {H.}~\bibnamefont
  {Ishida}}\ and\ \bibinfo {author} {\bibfnamefont {A.}~\bibnamefont
  {Liebsch}},\ }\Doi {10.1103/PhysRevB.82.045107} {\bibfield  {journal}
  {\bibinfo  {journal} {Phys. Rev. B},\ }\textbf {\bibinfo {volume} {82}},\
  \bibinfo {pages} {045107} (\bibinfo {year} {2010})}\BibitemShut {NoStop}%
\bibitem [{\citenamefont {Nourafkan}\ \emph {et~al.}(2009)\citenamefont
  {Nourafkan}, \citenamefont {Capone},\ and\ \citenamefont
  {Nafari}}]{Nourafkan2009}%
  \BibitemOpen
  \bibfield  {author} {\bibinfo {author} {\bibfnamefont {R.}~\bibnamefont
  {Nourafkan}}, \bibinfo {author} {\bibfnamefont {M.}~\bibnamefont {Capone}}, \
  and\ \bibinfo {author} {\bibfnamefont {N.}~\bibnamefont {Nafari}},\ }\Doi
  {10.1103/PhysRevB.80.155130} {\bibfield  {journal} {\bibinfo  {journal}
  {Phys. Rev. B},\ }\textbf {\bibinfo {volume} {80}},\ \bibinfo {pages}
  {155130} (\bibinfo {year} {2009})}\BibitemShut {NoStop}%
\bibitem [{\citenamefont {Kalkenstein}\ and\ \citenamefont
  {Soven}(1971)}]{Kalkenstein1971}%
  \BibitemOpen
  \bibfield  {author} {\bibinfo {author} {\bibfnamefont {D.}~\bibnamefont
  {Kalkenstein}}\ and\ \bibinfo {author} {\bibfnamefont {P.}~\bibnamefont
  {Soven}},\ }\Doi {10.1103/PhysRevB.82.115127} {\bibfield  {journal} {\bibinfo
   {journal} {Surf. Sci.},\ }\textbf {\bibinfo {volume} {26}},\ \bibinfo
  {pages} {85} (\bibinfo {year} {1971})}\BibitemShut {NoStop}%
\end{thebibliography}%
\end{document}